\documentclass[useAMS,usenatbib]{mnras}

\usepackage{myaasmacros}
\usepackage{graphicx}
\usepackage{ulem}
\usepackage[table]{xcolor}
\usepackage{amsmath}
\usepackage{amssymb}

\def\ltsima{$\; \buildrel < \over \sim \;$}
\def\simlt{\lower.5ex\hbox{\ltsima}}
\def\gtsima{$\; \buildrel > \over \sim \;$}
\def\simgt{\lower.5ex\hbox{\gtsima}}
\DeclareFontFamily{U}{mathx}{\hyphenchar\font45}
\DeclareFontShape{U}{mathx}{m}{n}{<-> mathx10}{}
\DeclareSymbolFont{mathx}{U}{mathx}{m}{n}
\DeclareMathAccent{\widebar}{0}{mathx}{"73}

\def\siglos{\sigma_{\text{LOS}}}
\def\siglosi{\sigma_{\text{LOS},i}}
\def\siglosobs{\overline{\sigma}_{\text{LOS}}}
\def\siglosobsi{\overline{\sigma}_{\text{LOS},i}}

\def\DM{\text{D}}
\def\betastar{\tilde{\beta}}
\def\dd{\text{d}}

\def\EMCEE{{\sc emcee}}

\def\sigpmr{\sigma_{\text{pmr}}}
\def\sigpmt{\sigma_{\text{pmt}}}

\def\sigpmrobsi{\overline{\sigma}_{\text{pmr},i}}
\def\sigpmtobsi{\overline{\sigma}_{\text{pmt},i}}
\def\Sigmai{\Sigma_{*,i}}
\def\Sigmaij{\Sigma_{*,ij}}
\def\Sigmaobs{\overline{\Sigma}_{*}}
\def\Sigmaobsi{\overline{\Sigma}_{*,i}}
\def\Sigmaobsij{\overline{\Sigma}_{*,ij}}
\def\vlosfour{\langle v_{\rm LOS}^4 \rangle}
\def\vlosfourobs{\overline{\langle v_{\rm LOS}^4 \rangle}}
\def\vsone{v_{s1}}
\def\vstwo{v_{s2}}
\def\vsoneobs{\overline{v}_{s1}}
\def\vstwoobs{\overline{v}_{s2}}
\def\vsoneobsi{\overline{v}_{s1,i}}
\def\vstwoobsi{\overline{v}_{s2,i}}

\def\GravSphere{{\sc GravSphere}}

\def\python{{\sc Python}}
\def\LMFIT{{\sc LMFIT}}
\def\GC{{\sc Gaia Challenge}}

\title[Breaking the density-anisotropy degeneracy in spherical systems]{How to break the density-anisotropy degeneracy in spherical stellar systems}

\author[Read]{J. I. Read$^1$\thanks{E-mail: justin.inglis.read@gmail.com}, P. Steger$^2$\\
  $^1$Department of Physics, University of Surrey, Guildford, GU2 7XH, UK\\
  $^2$Institute for Astronomy, Department of Physics, ETH Z\"urich, Wolfgang-Pauli-Strasse 27, CH-8093 Z\"urich, Switzerland}
  
\begin{document}

\maketitle

\begin{abstract}
We present a new non-parametric Jeans code, \GravSphere, that recovers the density $\rho(r)$ and velocity anisotropy $\beta(r)$ of spherical stellar systems, assuming only that they are in a steady-state. Using a large suite of mock data, we confirm that with only line-of-sight velocity data, \GravSphere\ provides a good estimate of the density at the projected stellar half mass radius, $\rho(R_{1/2})$, but is not able to measure $\rho(r)$ or $\beta(r)$, even with 10,000 tracer stars. We then test three popular methods for breaking this $\rho-\beta$ degeneracy: using multiple populations with different $R_{1/2}$; using higher order `Virial Shape Parameters' (VSPs); and including proper motion data. 

We find that two populations provide an excellent recovery of $\rho(r)$ in-between their respective $R_{1/2}$. However, even with a total of $\sim 7,000$ tracers, we are not able to well-constrain $\beta(r)$ for either population. By contrast, using 1000 tracers with higher order VSPs we are able to measure $\rho(r)$ over the range $0.5 < r/R_{1/2} < 2$ and broadly constrain $\beta(r)$. Including proper motion data for all stars gives an even better performance, with $\rho$ and $\beta$ well-measured over the range $0.25 < r/R_{1/2} < 4$.

Finally, we test \GravSphere\ on a triaxial mock galaxy that has axis ratios typical of a merger remnant, $[1:0.8:0.6]$. In this case, \GravSphere\ can become slightly biased. However, we find that when this occurs the data are poorly fit, allowing us to detect when such departures from spherical symmetry become problematic.
\end{abstract}


\section{Introduction}\label{sec:intro}

The mass modelling of stellar systems has a long and rich history dating back to \citet{1915MNRAS..75..366E}'s work on modelling star clusters and \citet{1922ApJ....55..302K}'s seminal model of the Milky Way. It has been used to probe the distribution and nature of dark matter in galaxies and galaxy clusters \citep[e.g.][]{1937ApJ....86..217Z,KleynaEtal2001,2003MNRAS.343..401L,2009AJ....137.3100W,2016ARA&A..54..597C}; the initial mass function of stars \citep[e.g.][]{2012Natur.484..485C,2012ApJ...755..156S,2014MNRAS.439..659N,2015MNRAS.448L..94S}; the structure of star clusters \citep[e.g.][]{1993ASPC...50..357P,2013MNRAS.436.2598W,2016MNRAS.462.2333P}; the mass of central `supermassive' black holes \citep[e.g.][]{MagorrianEtal1998,2010MNRAS.401.1770V}; and even the nature of weak-field gravity \citep[e.g.][]{2012LRR....15...10F,2014MNRAS.441.2497L,2016ApJ...832L...8M}.

To a very good approximation, stars in galaxies obey the collisionless Boltzmann equation \citep[e.g.][]{1987gady.book.....B}:

\begin{equation}
  \frac{\dd f}{\dd t} = \frac{\partial f}{\partial t} + \nabla_{\mathbf{x}} f\cdot\mathbf{v} - \nabla_{\mathbf{v}} f\cdot\nabla_{\mathbf{x}}\Phi = 0,
  \label{eqn:colboltz}
\end{equation}
where $f(\mathbf{x},\mathbf{v})$ is the distribution function of stars in phase space; $\mathbf{x}$ and $\mathbf{v}$ are the position and velocity of the stars, respectively; and $\Phi$ is the gravitational potential.

For galaxies in a `steady state', the $\frac{\partial f}{\partial t}$ term is small and can be discarded. Thus, in principle, given the positions and velocities of many stars, we can directly solve equation \ref{eqn:colboltz} for the force field $\nabla_{\mathbf{x}}\Phi$. From this we can derive the mass distribution from the Poisson equation: 

\begin{equation} 
\nabla_{\mathbf{x}}^2 \Phi = 4 \pi G \rho
\end{equation}
where $G = 6.67398\cdot10^{-11} \text{m}^3/\text{kg}\,\text{s}^2$ is Newton's gravitational constant, and the matter density $\rho$ includes all contributions from stars, gas and dark matter. However, in practice solving equation \ref{eqn:colboltz} is hard because it involves derivatives of $f$ that is six dimensional and poorly sampled. Several solutions have been proposed for this, including assuming a well-motivated functional form for $f$ \citep[e.g.][]{1915MNRAS..75..366E,1966AJ.....71...64K,1990ApJ...356..359H,1991MNRAS.253..414C,1995MNRAS.272..317H,2015MNRAS.454..576G}; modelling $f$ as a sum of many individual stellar orbits \citep[e.g.][]{1979ApJ...232..236S,2008MNRAS.385..647V,2013A&A...558A..35B}; or dynamically evolving an $N$-body simulation of $f$ towards a target distribution \citep[e.g.][]{1996MNRAS.282..223S,2007MNRAS.376...71D,2009MNRAS.395.1079D,2013MNRAS.430.1928H}. In this paper, we focus on a fourth approach: moment methods. These were first proposed by \citet{1922MNRAS..82..122J} and involve integrating equation \ref{eqn:colboltz} over each component of the velocity to create a set of three moment equations. There are several well-known disadvantages to this approach \citep[e.g.][]{1987gady.book.....B}: the hierarchy of moment equations is not closed; there is no guarantee that their solution will correspond to a physical distribution function; and the method typically requires the data to be binned, leading to the loss of potentially useful information. However, the key advantages are that the form of $f$ need not be specified and that the Jeans equations can be solved very rapidly. This latter is particularly important for the method we present in this work, where we employ a large number of free parameters that we marginalise over when fitting data.

In this paper, we concern ourselves with non-rotating near-spherical stellar systems that are a reasonable approximation to star clusters, spheroidal and elliptical galaxies and galaxy clusters. Integrating equation \ref{eqn:colboltz} over the radial velocity $v_r$ and assuming no rotation, we arrive at the `spherical Jeans equation' \citep{1980MNRAS.190..873B,1987gady.book.....B}: 

\begin{equation}
    \frac{1}{\nu}\frac{\partial}{\partial r}\left(\nu\sigma_{r}^2\right) +
    \frac{2\beta(r)\sigma_{r}^2}{r} = -\frac{GM(<r)}{r^2},\\
    \label{eqn:spherical_jeans}
\end{equation}
where:
\begin{equation}
\sigma_r^2 = \langle v_r^2 \rangle - \langle v_r \rangle^2 \,\,\,\, ; \,\,\,\, \langle v_r^n \rangle = \int v_r^n f d^3{\bf v};  
\label{eqn:moments}
\end{equation}
$M(<r)$ is the total cumulative mass as a function of radius $r$; $\nu$ and $\beta$ are the `tracer density' and `velocity anisotropy', respectively:

\begin{equation}
 \nu = \int f d^3{\bf v} \,\,\,\, ; \,\,\,\, \beta \equiv 1 - \frac{\sigma_t^2}{\sigma_r^2} 
 \label{eqn:nubeta}
\end{equation} 
and $\sigma_r$ and $\sigma_t$ are the radial and tangential velocity dispersion, respectively. 

The tracer density $\nu(r)$ describes the radial density profile of a population of massless tracers moving in the gravitational potential of a spherical mass distribution $M(r)$. These could be massless `stars' moving in a star cluster or galaxy, or massless galaxies moving in a galaxy cluster. If the tracers correspond to stars/galaxies that do have significant gravitating mass, then this is subsumed into the definition of $M(r)$. The velocity anisotropy $\beta(r)$ describes the orbital structure of the stellar system, where $\beta = 0$ corresponds to an isotropic distribution; $\beta = 1$ is a fully radial distribution; and $\beta = -\infty$ is a fully tangential distribution. 

Integrating both sides of equation \ref{eqn:spherical_jeans} gives the radial velocity dispersion as function of radius $r$ \citep[e.g.][]{1994MNRAS.270..271V,2005MNRAS.363..705M}:

\begin{equation}
\sigma_r^2(r) = \frac{1}{\nu(r) g(r)} \int_r^\infty \frac{GM(\tilde{r})\nu(\tilde{r})}{\tilde{r}^2} g(\tilde{r}) \dd \tilde{r}
\label{eqn:main}
\end{equation}
where:
\begin{equation}
g(r) = \exp\left(2\int \frac{\beta(r)}{r}\dd r\right)
\label{eqn:ffunc}
\end{equation}

However, typically, only line-of-sight velocities are observable. Projecting equation \ref{eqn:main} along the line of sight, we obtain \citep{1982MNRAS.200..361B}:

\begin{equation}
    \siglos^2(R) = \frac{2}{\Sigma_*(R)}\int_R^\infty \left(1-\beta\frac{R^2}{r^2}\right)
    \frac{\nu(r)\sigma_r^2(r) r}{\sqrt{r^2-R^2}}\dd r,
    \label{eqn:LOS}
\end{equation}
where $\Sigma_*(R)$ denotes the tracer surface mass density at projected radius $R$; and $\siglos(R)$ is the line-of-sight velocity dispersion of these tracer stars.

From equation \ref{eqn:LOS}, it is clear that the velocity anisotropy $\beta(r)$ trivially degenerates with $\sigma_r(r)$ and, therefore, with the cumulative mass distribution $M(<r)$ that we would like to measure. This is the well-known $M-\beta$ degeneracy \citep[e.g.][]{1990AJ.....99.1548M,2002MNRAS.330..778W,2003MNRAS.343..401L,2009MNRAS.395...76D}. Since the radial density profile $\rho(r)$ uniquely maps to the enclosed mass $M(<r)$, this can also be cast as an equivalent density-velocity anisotropy degeneracy that we refer to from here on as the `$\rho-\beta$ degeneracy'.

Several methods have been proposed in the literature to break the $\rho-\beta$ degeneracy. \citet{2000AJ....119..153S} suggest using multiple tracer populations, typically split by metallicity, colour, or stellar type (and see also \citealt{2008ApJ...681L..13B,2011ApJ...742...20W,2011MNRAS.tmp.1606A,2014MNRAS.439..659N}). \citet{2009MNRAS.399..812W} suggest performing a joint analysis on similar galaxy clusters to recover a statistical estimate of $\beta(r)$ for the whole population. \citet{1993ApJ...407..525V} suggest measuring the shape of the line-of-sight velocity distribution function, using deviations from Gaussianity to constrain $\beta(r)$ (and see also \citealt{2002MNRAS.330..778W,2013MNRAS.433.3173B,2013MNRAS.429.3079M}). \citet{2003MNRAS.343..401L} propose using higher order moments of $f$ -- e.g. the kurtosis -- to break the degeneracy. However, this can only work if assumptions are made about the form of $\beta(r)$ and its higher order counterparts \citep{1990AJ.....99.1548M,2013MNRAS.432.3361R}. For this reason, \citet{1990AJ.....99.1548M} propose instead using fourth order `Virial Shape Parameters' (VSPs; and see \citealt{2014MNRAS.441.1584R}). These are two fourth order moments integrated over radius $r$ to produce two scalar numbers (see \S\ref{sec:method}). The advantage of these VSPs is that they depend only on $\beta(r)$. Finally, proper motion data can be used to break the degeneracy \citep[e.g.][]{2007ApJ...657L...1S,2010ApJ...710.1063V,2013MNRAS.436.2598W}. 

While there is no shortage of methods for breaking the $\rho-\beta$ degeneracy, there is often a wide dispersion of conflicting results in the literature. For example, for the Sculptor dwarf spheroidal galaxy, split population methods favour a constant density dark matter core \citep{2008ApJ...681L..13B,2011ApJ...742...20W,2011MNRAS.tmp.1606A,2012ApJ...754L..39A}; a VSP method favours a cusp \citep{2014MNRAS.441.1584R}; and \citet{2013MNRAS.433.3173B} have argued, using a Schwarzschild distribution function method, that the data are not sufficiently constraining to be able to tell (and see \citealt{2014ApJ...791L...3B}). Similarly, there is an ongoing debate about the presence or absence of an `intermediate' mass black hole in the globular clusters Omega-Centauri \citep[e.g.][]{2008ApJ...676.1008N,2010ApJ...710.1063V,2010ApJ...719L..60N,2013MNRAS.436.2598W,2016IAUS..312..197Z} and G1 \citep[e.g.][]{2003ApJ...589L..25B,2005ApJ...634.1093G,2012ApJ...755L...1M}, while conflicting results are also found in the elliptical galaxy modelling community \citep[e.g.][]{2003Sci...301.1696R,2009MNRAS.395...76D}.

To shed light on the above discrepancies, in this paper we develop a new non-parametric spherical Jeans code, \GravSphere, that is designed to recover $\beta(r)$ and $\rho(r)$ under the assumptions only of spherical symmetry and a steady-state distribution function. We test \GravSphere\ on a large suite of spherical and triaxial mock data, probing different tracer distributions, $\beta(r)$ profiles, and $\rho(r)$ profiles. Our specific goal is to assess -- for ideal data -- how well different methods for breaking the $\rho-\beta$ degeneracy work. We focus in particular on three methods: split populations; VSPs; and proper motions. In a companion paper, we will describe these mock data in more detail and compare and contrast \GravSphere\ with over methods in the literature (Read et al. in prep.).

This paper is organised as follows. In \S\ref{sec:method}, we describe the \GravSphere\ method. In \S\ref{sec:mocks}, we briefly describe the mock data that we use in this paper that is a subset of the `default' spherical and triaxial suite from the \GC\ wiki site \citep{2011ApJ...742...20W,2009MNRAS.395.1079D}\footnote{\href{http://astrowiki.ph.surrey.ac.uk/dokuwiki/}{http://astrowiki.ph.surrey.ac.uk/dokuwiki/}.}. These mock data are publicly available and described in detail in Read et al. in prep. In \S\ref{sec:results}, we test \GravSphere\ on these mock data, assuming only line-of-sight data (\S\ref{sec:losonly}); including split populations (\S\ref{sec:split}); testing the utility of VSPs (\S\ref{sec:vs}); and adding in proper motion data (\S\ref{sec:proper}). In \S\ref{sec:triaxial}, we assess how well \GravSphere\ performs on triaxial data for which it should become biased. Finally, in \S\ref{sec:conclusions} we present our conclusions.

\section{Method}\label{sec:method}

In this section, we describe our new \GravSphere\ `non-parametric' Jeans modelling code. \GravSphere\ primarily solves equation \ref{eqn:LOS} to calculate the projected velocity dispersion profile, $\siglos(R)$, for a set of tracers moving within a spherical mass distribution $M(r)$. These `tracers' can be stars in star clusters or galaxies; globular clusters or planetary nebulae, or even galaxies moving in a galaxy cluster; they do not need to self consistently generate the full gravitational potential. At minimum, we assume that we have measurements of $\siglos(R)$ and the surface brightness profile of the tracers, $\Sigma_*(R)$. We also explore having $N_t$ distinct tracers that each have their own $\siglosi(R), \Sigmai(R)$ and velocity anisotropy $\beta_i(r)$, while moving in the same mass distribution $M(r)$. And we consider the utility of higher order `Virial Shape Parameters' (VSPs) and proper motion data. Otherwise, our goal is to assume only spherical symmetry and a steady-state distribution function. We describe our method as `non-parametric' because we have sufficient freedom in our functional forms for $\nu(r)$ and $M(r)$ that we can have many more parameters than data constraints. These are then marginalised over to explore model degeneracies (see e.g. the discussion in \citet{2014MNRAS.445.2181C} on this point). 

In the following subsections, we describe our functional forms for $\beta(r)$ (\S\ref{sec:betamodel}), $\nu(r)$ (\S\ref{sec:numodel}) and $M(r)$ (\S\ref{sec:Mmodel}); how we wrap in additional data from multiple populations (\S\ref{sec:multimethod}), VSPs (\S\ref{sec:VSPsmethod}) and proper motions (\S\ref{sec:propermethod}); and how we compare our models to data (\S\ref{sec:mcmcmethod}).

\subsection{The velocity anisotropy}\label{sec:betamodel}

We assume an expression for the velocity anisotropy $\beta(r)$ that ensures broad generality while making the function $g(r)$ (equation \ref{eqn:ffunc}) analytic: 

\begin{equation} 
\beta(r) = \beta_0 + \left(\beta_\infty-\beta_0\right)\frac{1}{1 + \left(\frac{r_0}{r}\right)^n}
\label{eqn:beta}
\end{equation}
where $\beta_0$ is the inner asymptotic anisotropy (in the limit $r\rightarrow 0$); $\beta_\infty$ is the outer asymptotic anisotropy (in the limit $r\rightarrow \infty$); $r_0$ is a transition radius; and $n$ controls the sharpness of the transition. Notice that equation \ref{eqn:beta} is identical to the Osipkov-Merritt anisotropy profile \citep{1979PAZh....5...77O,1985MNRAS.214P..25M}:

\begin{equation}
\beta = r^2 / (r^2 + r_0^2)
\label{eqn:betaosip}
\end{equation}
in the limit $n=2$, $\beta_0 = 0$ and $\beta_\infty = 1$.

Putting equation \ref{eqn:beta} into equation \ref{eqn:ffunc}, we derive:

\begin{equation} 
g(r) = r^{2\beta_\infty}\left(\left(\frac{r_0}{r}\right)^n+1\right)^{\frac{2}{n}(\beta_\infty-\beta_0)}
\label{eqn:betaf}
\end{equation} 
More general forms for $\beta(r)$ can be readily implemented if the data quality warrant it\footnote{In an early version of \GravSphere, we employed a fully non-parametric model for $\beta(r)$ that was linearly interpolated over a number of discrete bins. This was problematic for two reasons. Firstly, it required us to numerically solve the integral in equation  \ref{eqn:ffunc}. This must be solved with care since it is then used as the exponent of an exponential. Secondly, without regularisation priors, this led to a wildly fluctuating $\beta(r)$ that varied discretely from bin to bin. For these reasons, we moved instead to a smooth analytic form for $\beta(r)$.}. Indeed, the above form can be generalised to have $N$ radii at which $\beta(r)$ transitions to a different value: 

\begin{eqnarray}
\beta(r) = \beta_0 + (\beta_\infty-\beta_0)\left[\sum_{i=0}^{N}\left(1+\left(\frac{r_i}{r}\right)^{n_i}\right)^{-1}\right]
\end{eqnarray}
which also gives an analytic solution to equation \ref{eqn:ffunc}:
\begin{eqnarray}
g(r) & = & \exp\left[2\beta_0 \ln r + 2(\beta_\infty - \beta_0)\sum_{i=0}^N \frac{\ln\left(1+\left(\frac{r_i}{r}\right)^{n_i}\right)}{n_i}\right. + \nonumber \\
&& \left.2(\beta_\infty-\beta_0)\sum_{i=0}^N\ln r\right]
\end{eqnarray}
In this paper, we explore only the simpler form for $\beta$ in equation \ref{eqn:beta}.

Equation \ref{eqn:beta} is a useful form for the velocity anisotropy because it appears directly in equation \ref{eqn:LOS} and it avoids us needing to numerically integrate equation \ref{eqn:ffunc}. However, it is problematic for model fitting because $\beta_0$ and $\beta_\infty$ are defined over an infinite range ($-\infty < \beta_0 < 1$). For this reason, throughout this paper we work instead with a symmetrised anisotropy parameter \citep{2006MNRAS.tmp..153R}: 

\begin{equation} 
\betastar =  \frac{\sigma_r - \sigma_t}{\sigma_r + \sigma_t} = \frac{\beta}{2-\beta}
\label{eqn:betastar}
\end{equation} 
where $\betastar = -1$ corresponds to full tangential anisotropy; $\betastar = 1$ to full radial anisotropy; and $\betastar = 0$ to isotropy. By using $\betastar$ instead of $\beta$, we can set flat priors on $-1 < \betastar_{0,\infty} < 1$ that give equal weight to tangentially and radially anisotropic models. (The transformation from $\betastar_{0,\infty}$ to $\beta_{0,\infty}$ is given by equation \ref{eqn:betastar}.)

\subsection{The tracer density profile}\label{sec:numodel}

For the tracer density, we use a sum of $N_P$ Plummer spheres, similarly to \citet{2016MNRAS.459.3349R}: 

\begin{equation}
\nu = \sum_j^{N_P} \frac{3 M_j}{4\pi a_j^3}\left(1+\frac{r^2}{a_j^2}\right)^{-5/2}
\label{eqn:multiplumden}
\end{equation}
where $M_j$ and $a_j$ are the mass and scale length of each individual component. (These components should not be thought of as physical, but rather as terms in a density expansion.)

This has the advantage that the tracer surface density (that we will compare with observational data (\S\ref{sec:mcmcmethod}) and that appears in equation \ref{eqn:LOS}) is then analytic: 

\begin{equation}
\Sigma_*(R) =  \frac{R^2}{\rho(R)}\sum_j^{N_P} \frac{15 M_j}{4\pi a_j^5}\left(1+\frac{R^2}{a_j^2}\right)^{-7/2}
\label{eqn:multiplumsurf}
\end{equation} 
We fit the above expansion {\it in advance} of running our full fit using the \python\ package \LMFIT\ \citep{newville_2014_11813}\footnote{\href{http://zenodo.org/record/11813\#.V2P8khV97eQ}{http://zenodo.org/record/11813\#.V2P8khV97eQ}.}. We then allow $M_j$ and $a_j$ to vary in the final fits by up to 50\% of their best-fit \LMFIT\ values. This makes the method substantially more efficient since $\Sigma_*(R)$ is usually much better constrained than the stellar kinematics. It also avoids sampling trivial degeneracies between the individual Plummer components in equation \ref{eqn:multiplumden}. By default, we use ${N_P}=3$ Plummer components, corresponding to $6$ free parameters, but arbitrarily many can be employed if the data require it. (Note that the above expansion is similar to the Multi-Gaussian Expansion proposed by \citet{1994A&A...285..723E}. However, since both dwarf galaxies and globular star clusters are well-approximated by Plummer spheres, for many applications the above expansion is particularly efficient; e.g. \citealt{1911MNRAS..71..460P,1995MNRAS.277.1354I,2008ApJ...684.1075M}.)

\subsection{The mass profile}\label{sec:Mmodel}

We split the enclosed mass profile into a contribution from stars, $M_*$, and a contribution from dark matter $M_{\rm dm}$:

\begin{equation}
M(r) = M_*(r) + M_{\rm dm}(r)
\end{equation}
The stellar component is assumed to have the same form as the photometric light profile (fit similarly to the tracer density, as described in \S\ref{sec:numodel}) with a mass to light ratio parameter $\Upsilon$ that we fit for. For the mock data in this paper, $\Upsilon = 0$ and the stars do not contribute to the gravitational potential (see \S\ref{sec:mocks}). 

The dark matter component is described by a sequence of power laws defined in $N_{\rm dm}$ radial bins with bin edges $r_j$: 

\begin{equation}
\rho_{\rm dm}(r) = \left\{
\begin{array}{ll}
\rho_0\left(\frac{r}{r_0}\right)^{-\gamma_0} & r < r_0 \\ 
\rho_0\displaystyle\prod_{n=0}^{j<N_{\rm dm}}\left(\frac{r_{n+1}}{r_n}\right)^{-\gamma_n}\left(\frac{r}{r_{j+1}}\right)^{-\gamma_{j+1}} & r_j < r < r_{j+1} \\ 
\label{eqn:rhodm}
\end{array}
\right.
\end{equation}
where $\rho_0$ is a normalisation parameter that controls the mass of the dark matter halo, and $\gamma_j$ sets the logarithmic density slope of each bin $j$. The cumulative mass profile $M_{\rm dm}(r)$ then follows analytically from integrating equation \ref{eqn:rhodm}.

The above form allows a wide range of density profiles to be explored while ensuring that these are positive definite\footnote{This is important, for example, in satellite dwarf galaxies that may have their inner dark matter `heated' by bursty star formation \citep[e.g.][]{1996MNRAS.283L..72N,2005MNRAS.356..107R,2012MNRAS.421.3464P} and their outer dark matter profile sculpted by tides \citep[e.g.][]{2006MNRAS.tmp..153R}. These two processes introduce two new scales to the dark matter halo on which we expect it to deviate from the `NFW' profile found in $\Lambda$ Cold Dark Matter  `dark matter only' structure formation simulations \citep{1996ApJ...462..563N}.}. It also defines the dark matter density in terms of its local logarithmic slope, $\gamma_j$. This is a quantity of key interest for probing the nature of dark matter, but often difficult to extract because it requires a double differential of $M(<r)$ that can become very noisy. With the above functional form, we fit directly for $\gamma_j$, thereby avoiding this problem. Finally, we are able to place useful theory priors on $\gamma_j$ that it must be positive and monotonically decrease outwards. The former ensures that the density monotonically decreases outwards; the latter prevents the dark matter profile from having any inflection points. If such inflection points are expected in a given dark matter theory, then this prior can be relaxed \citep[e.g.][]{2016JCAP...05..010Y,2016arXiv160905856G}.

For a single tracer with only line-of-sight velocities, $M(<r)$ is maximally constrained at $\sim R_{1/2}$ \citep[e.g.][]{2009ApJ...704.1274W,2010MNRAS.406.1220W}. For this reason, we use as default $N_{\rm dm} = 5$ dark matter mass bins spaced logarithmically around $R_{1/2}$: 

\begin{equation}
r_{j=0 ... (N_{\rm dm}-1)} = [0.25, 0.5, 1.0, 2.0, 4.0]R_{1/2}
\end{equation}
More bins can be added if the data quality warrant it; we find that, for the mocks presented in this paper, we are not sensitive to this choice.

By default, for a single tracer component, our model has 6 parameters to describe the light profile (that are typically very well constrained by the data); 4 parameters to describe $\beta(r)$; 6 parameters to describe the mass profile ($\rho_0$ and 5 $\gamma_j$ for each radial bin); and one parameter to describe the mass to light ratio of the stars $\Upsilon$. This gives 17 parameters in total.

\subsection{Multiple tracer populations}\label{sec:multimethod}

Our method can cope naturally with $N_t$ sets of tracer stars moving in the same potential. As discussed in \S\ref{sec:intro}, if the $N_t$ populations have distinct $R_{1/2,i ... N_t}$, it has been argued in the literature that they can break the $\rho-\beta$ degeneracy and allow for a measurement of $\rho(r)$.

Each tracer $i$ has a distinct light profile $\nu_i(r)$ and velocity anisotropy profile $\beta_i(r)$. These lead to distinct projected light ($\Sigmai(R)$) and velocity dispersion ($\siglosi(R)$) profiles that can be compared with data. Thus, for each tracer we must solve an additional equation \ref{eqn:LOS}. 

Since there are now multiple $R_{1/2,i}$, we also update our binning strategy for $M_{\rm dm}$, placing one bin at each $R_{1/2,i}$, two at $[0.5,1]R_{1/2,{\rm min}}$ and two at $[2.0,4.0]R_{1/2,{\rm max}}$. Thus, our default method for $N_t$ tracers requires a total of $10N_t$ parameters to describe $\nu_i(r)$ and $\beta_i(r)$; and $6+N_t$ parameters to describe $M(r)$ (one for $\rho_0$; $4+N_t$ for the $\gamma_j$; and one for $\Upsilon$). This gives $11 N_t + 6$ parameters in total, which can become substantially more expensive than the single tracer case. 

\subsection{Including `Virial Shape Parameters' (VSPs)}\label{sec:VSPsmethod}

Following \citet{1990AJ.....99.1548M} and \citet{2014MNRAS.441.1584R}, we also include the ability to use higher order moments of the velocity distribution via the fourth order `Virial Shape Parameters' (VSPs):

\begin{eqnarray} 
\vsone & = & \frac{2}{5} \int_0^{\infty} GM \nu (5-2\beta) \sigma_r^2 r dr \\
\label{eqn:vs1}
& = & \int_0^{\infty} \Sigma_* \vlosfour R dR
\label{eqn:vs1data}
\end{eqnarray}
and
\begin{eqnarray} 
\vstwo & = & \frac{4}{35} \int_0^{\infty} GM \nu (7-6\beta) \sigma_r^2 r^3 dr \\
\label{eqn:vs2}
& = & \int_0^{\infty} \Sigma_* \vlosfour R^3 dR 
\label{eqn:vs2data}
\end{eqnarray}
The key advantage of these VSPs is that they involve fourth-order moments of the line of sight velocities $\vlosfour$ (equations \ref{eqn:vs1data} and \ref{eqn:vs2data}), but by integrating over projected radius $R$, $\vsone$ and $\vstwo$ depend only on $\beta$ and not on its fourth-order counterparts \citep[see equations \ref{eqn:vs1} and \ref{eqn:vs2} and][]{1990AJ.....99.1548M,2014MNRAS.441.1584R}. Thus, $\vsone$ and $\vstwo$ allow us to obtain constraints on $\beta$ via line of sight velocities alone, breaking the $\rho-\beta$ degeneracy. 

However, we take a slightly different approach to \citet{2014MNRAS.441.1584R} in comparing $\vsone$ and $\vstwo$ to data. They suggest recasting $\vsone$ and $\vstwo$ as dimensionless estimators: 

\begin{eqnarray} 
\zeta_A & = & N_{\rm tot} \frac{\int_0^{\infty} \Sigma_* \vlosfour R dR}{\left(\int_0^{\infty} \Sigma_*  \langle v_{{\rm LOS},i}^2 \rangle R dR\right)^2} \\
\label{eqn:zetaAtrue}
& \simeq & N_s \frac{\sum_i^{N_s} v_{{\rm LOS},i}^4}{\left(\sum_i^{N_s} v_{{\rm LOS},i}^2\right)^2}
\label{eqn:zetaA}
\end{eqnarray}  
\begin{eqnarray} 
\zeta_B & = & N_{\rm tot}^2 \frac{\int_0^{\infty} \Sigma_* \vlosfour R^3 dR}{\left(\int_0^{\infty} \Sigma_* \langle v_{\rm LOS}^2\rangle R dR\right)^2\left(\int_0^{\infty} \Sigma_* R^3 dR\right)^2} \\
\label{eqn:zetaBtrue}
& \simeq & N_s^2 \frac{\sum_i^{N_s} v_{{\rm LOS},i}^4 R_i^2}{\left(\sum_i^{N_s} v_{{\rm LOS},i}^2\right)^2\sum_i^{N_s} R_i^2}
\label{eqn:zetaB}
\end{eqnarray}
where $N_{\rm tot} = \int_0^{\infty} \Sigma_* R dR$ is $1/2\pi$ times the total number of stars, and $N_s$ is the actual number of observed stars.

The problem with $\zeta_A$ and $\zeta_B$ is twofold. Firstly, when calculating \ref{eqn:zetaAtrue} and \ref{eqn:zetaBtrue} from the data, we must take a noisy fourth order estimator and divide it by a noisy second order estimator, amplifying noise in the data. Secondly, the approximations in equations \ref{eqn:zetaA} and \ref{eqn:zetaB} allow for the inclusion only of errors in the velocity measurements. This could lead to artificially small errors on $\zeta_A$ and $\zeta_B$. For these reasons, in \GravSphere\ we work directly with $\vsone$ and $\vstwo$, solving equations \ref{eqn:vs1data} and \ref{eqn:vs2data} by numerically integrating the observed $\vlosfour$ over our best-fit model $\Sigma_*$. In this way, we can estimate the uncertainties on $\vsone$ and $\vstwo$ by Monte-Carlo sampling the errors on both $\vlosfour$ and, if significant, $\Sigma_*$. We describe this procedure in more detail in \ref{sec:mcmcmethod}. We compare our estimators for $\vsone$ and $\vstwo$ with those used in \citet{2014MNRAS.441.1584R} in Appendix \ref{app:estimators}.

\subsection{Including proper motions}\label{sec:propermethod}

Where proper motion data are available, we add two further constraint equations for the radially and tangentially averaged projected proper motions, $\sigma_{\rm pmr}$ and $\sigma_{\rm pmt}$ \citep[e.g.][]{2007ApJ...657L...1S,2010ApJ...710.1063V}: 

\begin{equation} 
\sigpmr(R) = \frac{2}{\Sigma_*(R)}\int_R^\infty \left(1-\beta+\beta\frac{R^2}{r^2}\right)
    \frac{\nu(r)\sigma_r^2(r) r}{\sqrt{r^2-R^2}}\dd r
    \label{eqn:sigpmr}
\end{equation}
\begin{equation} 
\sigpmt(R) = \frac{2}{\Sigma_*(R)}\int_R^\infty \left(1-\beta\right)
    \frac{\nu(r)\sigma_r^2(r) r}{\sqrt{r^2-R^2}}\dd r
    \label{eqn:sigpmt}
\end{equation}
Equations \ref{eqn:LOS}, \ref{eqn:sigpmr} and \ref{eqn:sigpmt} each have a different dependence on $\beta$. For an isotropic distribution ($\beta = 0$) all three become equal, while any anisotropy will cause them to differ. Thus, by fitting all three simultaneously to data for $\siglos$, $\sigpmr$ and $\sigpmt$, we can obtain tight constraints on both $\beta$ and $\rho$, thereby breaking the $\rho-\beta$ degeneracy (see \S\ref{sec:intro}).

\subsection{Fitting the model to data}\label{sec:mcmcmethod}

To compare our model to data, we bin the data in projected radius $R$. Following \citet{2009ApJ...704.1274W}, we use $N_{\rm bin} = \sqrt{N_*}$ bins where $N_*$ is the total number of stellar tracers. This gives us $\sqrt{N_*}$ stars in each bin, ensuring good sampling and equal Poisson error for each bin, while not compromising too much on the radial coverage. (In Appendix \ref{app:binning}, we show that we are not sensitive to this choice.) The error on each bin includes the effect of Poisson sampling and measurement error, calculated as in \citet{1993ASPC...50..357P}. From here on, we will denote binned data quantities with an line over the top. For example, the binned projected surface density of a tracer population $i$ will be denoted $\Sigmaobsi$ to differentiate it from the \GravSphere\ model surface density $\Sigmai$ in equation \ref{eqn:multiplumsurf}.

Our binned data comprise the projected surface density of each set of tracers $\Sigmaobsi$; the projected velocity dispersion of these tracers $\siglosobsi$ and, if proper motions are available, the radial and tangential projected dispersions $\sigpmrobsi$ and $\sigpmtobsi$. (In general, each of $\Sigmaobsi$, $\siglosobsi$, $\sigpmrobsi$ and $\sigpmtobsi$ will have its own bins.) 

To estimate the VSPs for a given tracer population, we use a novel technique. First, we calculate the binned fourth order line-of-sight velocity moment $\vlosfourobs$, using the same bins as for $\siglosobs$. We estimate the errors on $\vlosfourobs$ by Monte-Carlo sampling the velocity error on each star 1000 times and then adding the measurement variance on $\vlosfourobs$ in quadrature with the Poisson error. We then numerically integrate equations \ref{eqn:vs1} and \ref{eqn:vs2} to calculate $\vsoneobs$ and $\vstwoobs$. For these integrals, we use our best-fit model $\Sigma_*(R)$, assuming that the error on this is negligible as compared with the error on $\vlosfourobs$. To obtain an error on $\vsoneobs$ and $\vstwoobs$, we then perform a Monte-Carlo sampling of the errors on $\vlosfourobs$, repeating our numerical integration of equations \ref{eqn:vs1} and \ref{eqn:vs2} 1000 times. The variance on these integrals is then taken as a Gaussian error on $\vsoneobs$ and $\vstwoobs$. (We verified that the probability distribution functions of $\vsoneobs$ and $\vstwoobs$ that result from these integrals are indeed close to Gaussian and that our error bars are not sensitive to the choice of 1000 Monte-Carlo samples.)

We then fit our model to these binned data to obtain a joint $\chi^2$: 

\begin{eqnarray}
\chi^2 & = & \sum_i^{N_t} \sum_j^{N_{\rm bin}} \frac{(\Sigmaobsij - \Sigmaij)^2}{\sigma_{\Sigma_*,ij}^2} + ... \\ \nonumber 
\end{eqnarray} 
where $\Sigmaobsij$ is the observed surface density of a given tracer $i$ in bin $j$; $\Sigmaij$ is the model surface density of tracer $i$ evaluated at bin $j$; $\sigma_{\Sigma_*,ij}$ is the error on that bin; and similar $\chi^2$ terms are added for $\siglosobsi$, $\sigpmrobsi$ and $\sigpmtobsi$ (if available), and $\vsoneobsi$ and $\vstwoobsi$ (if available).

We fit our model to the data using the \EMCEE\ affine invariant Markov Chain Monte Carlo (MCMC) sampler from \citet{2013PASP..125..306F}. We assume uncorrelated Gaussian errors such that the Likelihood function is given by $\mathcal{L} = \exp(-\chi^2/2)$. We use as default 1000 walkers, each generating 5000 models and we throw out the first half of these as a conservative `burn in' criteria. (These numbers are substantially larger than usually employed due to the high dimension of our parameter space.) We explicitly checked that our results are converged by running more models and examining walker convergence. 

In setting our priors, our basic philosophy is to use flat priors on parameters that are well constrained and/or dimensionless, and logarithmic priors on everything else. Specifically, we use flat priors on the dark matter logarithmic slope parameters $0 < \gamma_j < 3$; a flat logarithmic prior on the dark matter density normalisation $6 < \log_{10}[\rho_0/({\rm M}_\odot\,{\rm kpc}^{-3})] < 10$; a flat prior on the stellar mass to light ratio $\Upsilon$ over a range $\pm 25$\% of the stellar population synthesis model value (for the mock data we use in this paper $\Upsilon = 0$); flat priors on the scale length parameters describing the tracer surface density $a_j$ within 50\% of their pre-determined best-fit values; a flat logarithmic prior on the mass parameters describing the tracer surface density $\log_{10}[M_j]$, also within 50\% of their pre-determined best-fit values; a flat logarithmic prior on the velocity anisotropy scale length $-0.3 < \log_{10}[r_0/R_{1/2}] < 0.3$; and a flat prior on the transition rate parameter $1 < n < 3$. We use priors on the inner and outer asymptotic velocity anisotropy parameters $\beta_0$ and $\beta_\infty$, as described in \S\ref{sec:betamodel}. Our results are not sensitive to these prior choices. Finally, we do not include any density slope-anisotropy prior. Such a prior could be used to ensure that the distribution function is everywhere positive \citep[e.g.][]{2006ApJ...642..752A,2010MNRAS.408.1070C}. We will explore this in future work.

\section{The mock data}\label{sec:mocks}

\begin{table*}
    \begin{center}
    \begin{tabular}{lccccc}
        Label & $[r_*,\alpha_*,\beta_*,\gamma_*]$ & $[(\rho_0,r_\DM,\alpha_\DM,\beta_\DM,\gamma_\DM)]$ & $\beta(r)$ & $R_{1/2}$ & $[a,b,c]$ \\
	\hline
	\hline
\rowcolor{gray!40} PlumCuspIso & $[0.25,2,5,0.1]$ & $[6.4,1,1,3,1]$ & 0 & 0.25 & $[1,1,1]$ \\
        NonPlumCoreOm & $[0.25,2,5,1]$ & $[40,1,1,3,0]$ & Eq. \ref{eqn:betaosip}; $r_a = 0.25$ & 0.20 & $[1,1,1]$ \\
\rowcolor{gray!40} NonPlumCuspOm & $[0.1,2,5,1]$ & $[6.4,1,1,3,1]$ & Eq. \ref{eqn:betaosip}; $r_a = 0.1$ & 0.079 & $[1,1,1]$ \\
PlumCuspTan & $[0.5,0.5,5,0.1]$ & $[2.39,2,1,4,1]$ & $-0.5$ & 0.86 & $[1,1,1]$ \\
\rowcolor{gray!40} SplitCompCore & $[0.5,2,5,1]_1$ & $[40,1,1,3,0]$ & Eq. \ref{eqn:betaosip}; $r_{a,1} = 0.5$ & 0.39 & $[1,1,1]$ \\
\rowcolor{gray!40} & $[1.0,2,5,1]_2$ & & Eq. \ref{eqn:betaosip}; $r_{a,2} = 1.0$ & 0.79 & \\
SplitCompCusp & $[0.5,2,5,1]_1$ & $[6.4,1,1,3,1]$ & Eq. \ref{eqn:betaosip}; $r_{a,1} = 0.5$ & 0.39 & $[1,1,1]$ \\
& $[1.0,2,5,1]_2$ & & Eq. \ref{eqn:betaosip}; $r_{a,2} = 1.0$ & 0.79 & \\
\rowcolor{gray!40} TriaxCusp & $[0.81,0.34,5.92,0.23]$ & $[5.5,1.5,1,4,1]$ & Eq. \ref{eqn:betatriax} & 0.6 ($X$), 0.8 ($Z$) & $[1,0.8,0.6]$ \\
        \hline
        \hline
    \end{tabular}
    \end{center}
    \caption{Parameters of the spherical and triaxial mock data that we use in this paper. The columns give, from left to right: the mock data label; the tracer density parameters (see equation \ref{eqn:zhao}); the dark matter halo density parameters (see equation \ref{eqn:zhao}); the velocity anisotropy profile; the projected half stellar mass radius; and the long $a$, intermediate $b$ and short $c$ axis of the figure, where $[1,1,1]$ corresponds to spherical symmetry. The length units are given in kpc, while $\rho_0$ has units of $10^7$\,M$_\odot$\,kpc$^{-3}$. The mock labelling convention is: $<$light profile$>$$<$Cusp/core$>$$<$Anisotropy$>$, where the $<$light profile$>$ can be Plummer-like (Plum; $\gamma_* = 0.1$); cusped (NonPlum; $\gamma_* = 1$); split-component (Split); or triaxial (Triax); the dark matter halo can be cusped (Cusp; $\gamma_\DM = 1$) or cored (Core; $\gamma_\DM = 0$); and the $<$Anisotropy$>$ can be isotropic (Iso; $\beta = 0$); of Osipkov-Merritt form (Om; see equation \ref{eqn:betaosip}) or tangential (Tan; $\beta = -0.5$). For the split-component models, there are two separate tracers moving in the same dark matter halo; these are labelled by the subscript 1 and 2, respectively. For the TriaxCusp mock, the stars and dark matter are assumed to be aligned, with identical axis ratios $[a,b,c]$. We project this mock along either the long axis ($X$) or short axis ($Z$), leading to two different $R_{1/2}$, as marked in the Table.}
    \label{tab:mocks}
\end{table*}

We take mock data from the `default spherical-triaxial suite' of the \GC\ wiki site\footnote{\href{http://astrowiki.ph.surrey.ac.uk/dokuwiki/}{http://astrowiki.ph.surrey.ac.uk/dokuwiki/}.}. The full spherical-triaxial suite is described in detail in Read et al. in prep. We have run all of the mock data in the default suite using \GravSphere. However, for brevity, we show results here for a representative subset of mocks as detailed in Table \ref{tab:mocks}. These are chosen to highlight both the best and worst performance of \GravSphere\ over all of the mock data.

The spherical mocks assume that both the tracer distribution and their host dark matter halo are given by a split power law form \citep{1996MNRAS.278..488Z}:
\begin{equation}
  \rho_X(r) = \rho_0\left(\frac{r}{r_X}\right)^{-\gamma_X} \left[1+\left(\frac{r}{r_X}\right)^{\alpha_X}\right]^{(\gamma_X-\beta_X)/\alpha_X}
  \label{eqn:zhao}
\end{equation}
where $\rho_0$ is a normalisation parameter; $r_X$ sets the scale length; $\gamma_X$ is the inner asymptotic logarithmic slope; $\beta_X$ is the outer asymptotic logarithmic slope; and $\alpha_X$ controls how sharp the transition is at $r_X$ between $\gamma_X$ and $\beta_X$. For the tracers, we use the notation: $\nu = \rho_*(r,\nu_0,r_*,\alpha_*,\beta_*,\gamma_*)$; for the dark matter we write similarly: $\rho_\DM = \rho_\DM(r,\rho_0,r_\DM,\alpha_\DM,\beta_\DM,\gamma_\DM)$.

For the radially anisotropic mocks, the tracers are set up in equilibrium inside their host dark matter halo assuming an Osipkov-Merritt distribution function \citep{1979PAZh....5...77O,1985MNRAS.214P..25M}, as in \citet{2011ApJ...742...20W}. This has an anisotropy profile $\beta(r)$ given by equation \ref{eqn:betaosip}, parameterised by the anisotropy radius $r_a$. The tangentially anisotropic mocks are set up using the `made to measure' code from \citet{2009MNRAS.395.1079D}. These also assume the form given in equation \ref{eqn:zhao} for the tracer and dark matter density profiles. The velocity anisotropy is, however, assumed to be a constant $\beta = -0.5$.

In addition to the above `single population' mocks, we consider two `split population' mocks that have two sets of tracers stars with distinct $R_{1/2}$ moving in the same dark matter halo. Like the single population mocks, these models assume an Osipkov-Merritt distribution function \citep{2011ApJ...742...20W}, where both tracers have a density profile of the form given in equation \ref{eqn:zhao}, but with different parameters and different numbers of tracer stars. We report the model parameters for these mocks in Table \ref{tab:mocks}. For this paper, we assume that we can perfectly split these two populations into their constituent stars. 

Finally, we consider a triaxial mock galaxy for which \GravSphere\ (that assumes spherical symmetry) should become biased. This was set up, similarly to the tangential mocks, using the \citet{2009MNRAS.395.1079D} made to measure code. As with the other mocks, the tracer stars and dark matter halo are assumed to be of the form given in equation \ref{eqn:zhao}. However, now the radius $r$ is replaced by an ellipsoidal radius parameter $m$: 

\begin{equation}
r^2 \rightarrow m^2 = \left(\frac{x}{a}\right)^2 + \left(\frac{y}{b}\right)^2 + \left(\frac{z}{c}\right)^2
\end{equation}
where $a > b > c$ set the long, intermediate and short axis of the triaxial figure. The mock we use here has $b/a=0.8$ and $c/a=0.6$, consistent with the typical triaxiality that occurs following the merger of two spherical galaxies \citep[e.g.][]{2017MNRAS.464.2301G}. The tracer stars and their host dark matter halo were assumed to have exactly the same triaxiality. The velocity anisotropy profile for the triaxial mocks is given by: 

\begin{equation} 
\beta(r) = \frac{r_t^{1/2}\beta_0 + r^{1/2}\beta_\infty}{r^{1/2} + r_t^{1/2}}
\label{eqn:betatriax}
\end{equation}
where $r_t = 0.81$\,kpc, $\beta_0 = 0$ and $\beta_\infty = 0.5$, such that these models are isotropic at $r=0$ and become radially anisotropic at radii $r > r_t$.

The parameters and labels that we use for all of our mock data are reported in Table \ref{tab:mocks}. In all mocks, the stars are {\it massless tracers} that do not contribute to the gravitational potential\footnote{For this reason, we do not report the value of $\nu_0$ in Table \ref{tab:mocks} since this parameter is arbitrary.}. For the single component mocks, we consider 1000 and 10,000 tracer stars; for the split component mocks, we use $[1877,4703]$ stars for the inner/outer tracer population in SplitCompCore, and $[3075,4030]$ for SplitCompCusp. In this way, the total number of stars for the split component mocks is comparable to the 10,000 star case for our single component tests. In all cases, we assume excellent data quality with constant errors of $2$\,km\,s$^{-1}$ on the measured velocities (line of sight or proper motion) for each star.

\section{Results}\label{sec:results}

\begin{figure*}
\begin{center}
\includegraphics[width=0.75\textwidth]{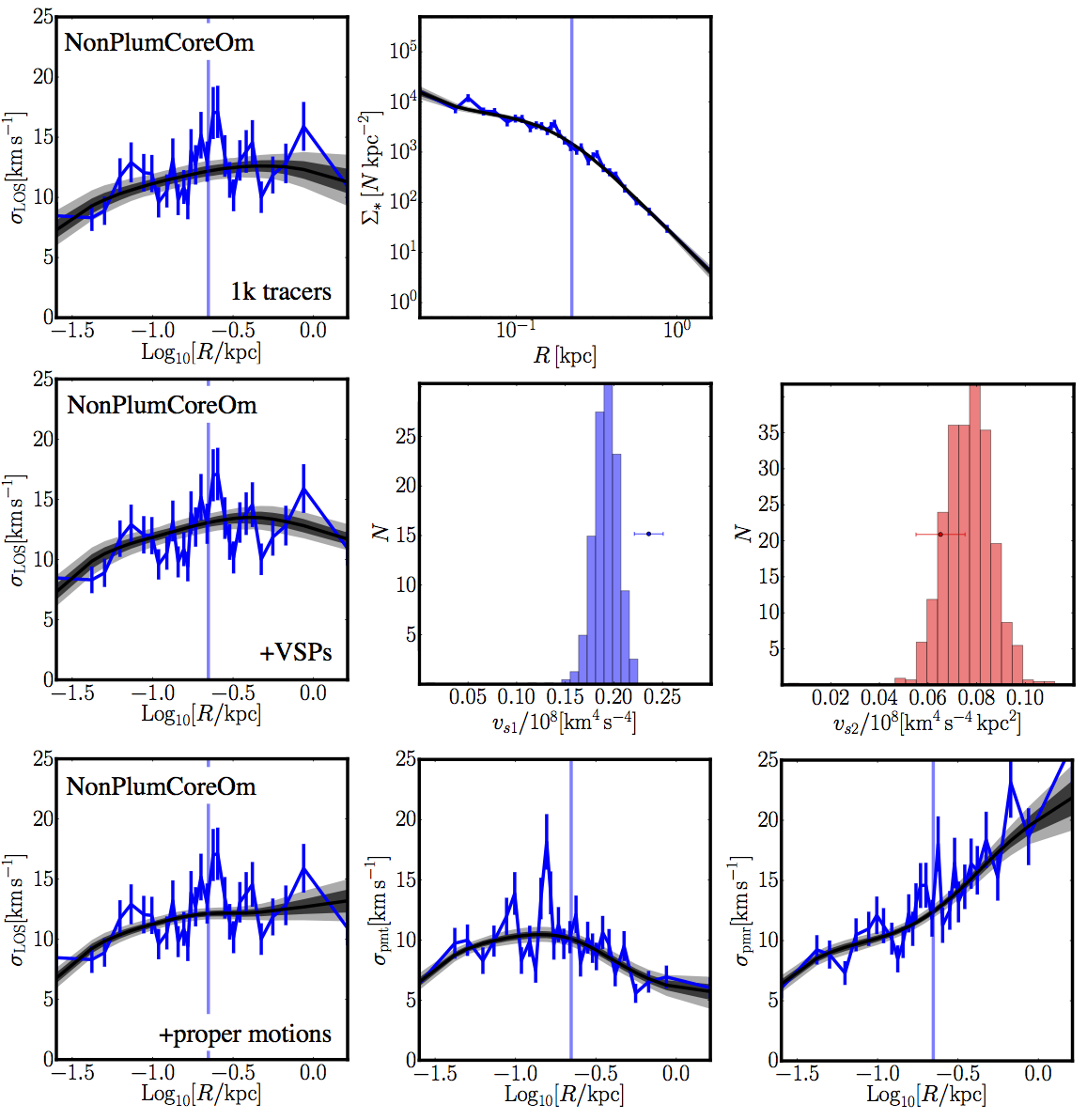}
\caption{Example \GravSphere\ fits for the NonPlumCoreOm mock data. {\bf Top row:} the median (black), 68\% (dark grey) and 95\% (light grey) confidence intervals of our fit for 1000 tracer stars, using only the line-of-sight velocities $\siglosobs(R)$ (left) and surface density of the tracer stars $\Sigmaobs(R)$ (right). The data are shown by the blue data points with error bars (that include the Poisson error and the measurement error for each star; see \S\ref{sec:mocks}). The vertical blue line marks the projected half stellar mass radius of the tracer stars, $R_{1/2}$. {\bf Middle row:} fits for the same as the top row but including the VSPs: $\vsone$ (middle) and $\vstwo$ (right; see \S\ref{sec:VSPsmethod} and \S\ref{sec:vs}). The two additional VSP data points that we fit are marked on these histograms, with errors calculated as in \S\ref{sec:mcmcmethod}. {\bf Bottom row:} fits for the same as the top row but including proper motions: the tangential projected dispersion $\sigpmt$ (middle) and the radial projected dispersion $\sigpmr$ (right; see \S\ref{sec:propermethod} and \S\ref{sec:proper}).}
\label{fig:FitExample} 
\end{center}
\end{figure*}

\begin{figure*}
\begin{center}
\includegraphics[width=0.75\textwidth]{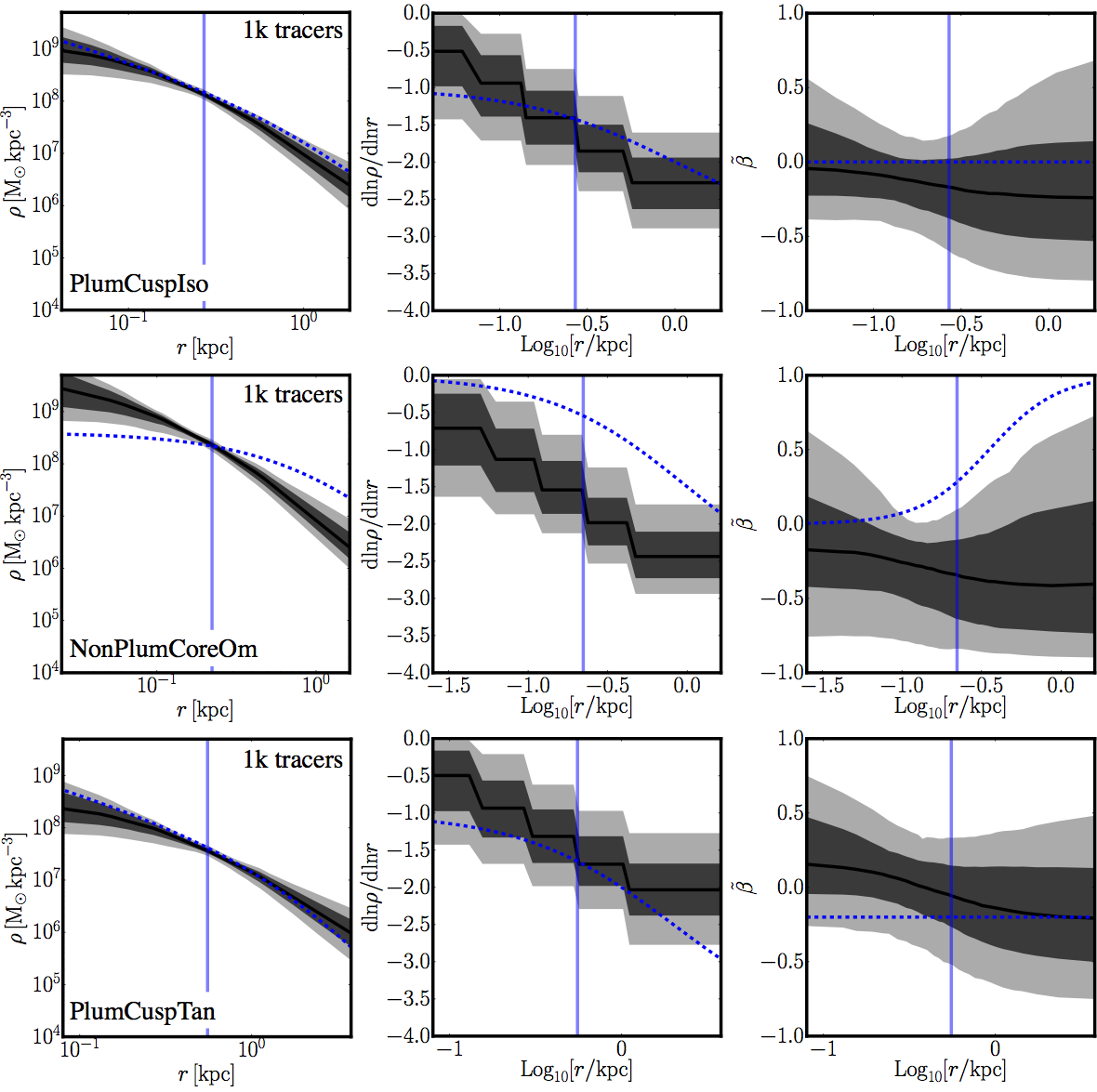}
\caption{Testing \GravSphere\ with only line-of-sight velocity data, $\siglosobs(R)$, and the surface density of the tracer stars $\Sigmaobs$. From left to right, the panels show the median (black), 68\% (dark grey) and 95\% (light grey) confidence intervals of the radial density profile $\rho(r)$; the logarithmic slope of the density profile $d\ln\rho/d\ln r$; and the symmetrised velocity anisotropy parameter $\betastar$ (equation \ref{eqn:betastar}). These tests used 1000 tracer stars. The correct answers are overlaid on each panel by the blue dashed lines. The vertical blue lines mark the projected half stellar mass radius of the tracer stars, $R_{1/2}$. From top to bottom, the panels show results for the spherical single component mocks: PlumCuspIso, NonPlumCoreOm and PlumCuspTan (see Table \ref{tab:mocks}). In Figure \ref{fig:FitExample}, top panels, we show how well $\siglosobs(R)$ and $\Sigmaobs$ are fit for the NonPlumCoreOm mock (\GravSphere\ performs similarly well for the other mocks).}
\label{fig:LOS_plots} 
\end{center}
\end{figure*}

\begin{figure*}
\begin{center}
\includegraphics[width=0.75\textwidth]{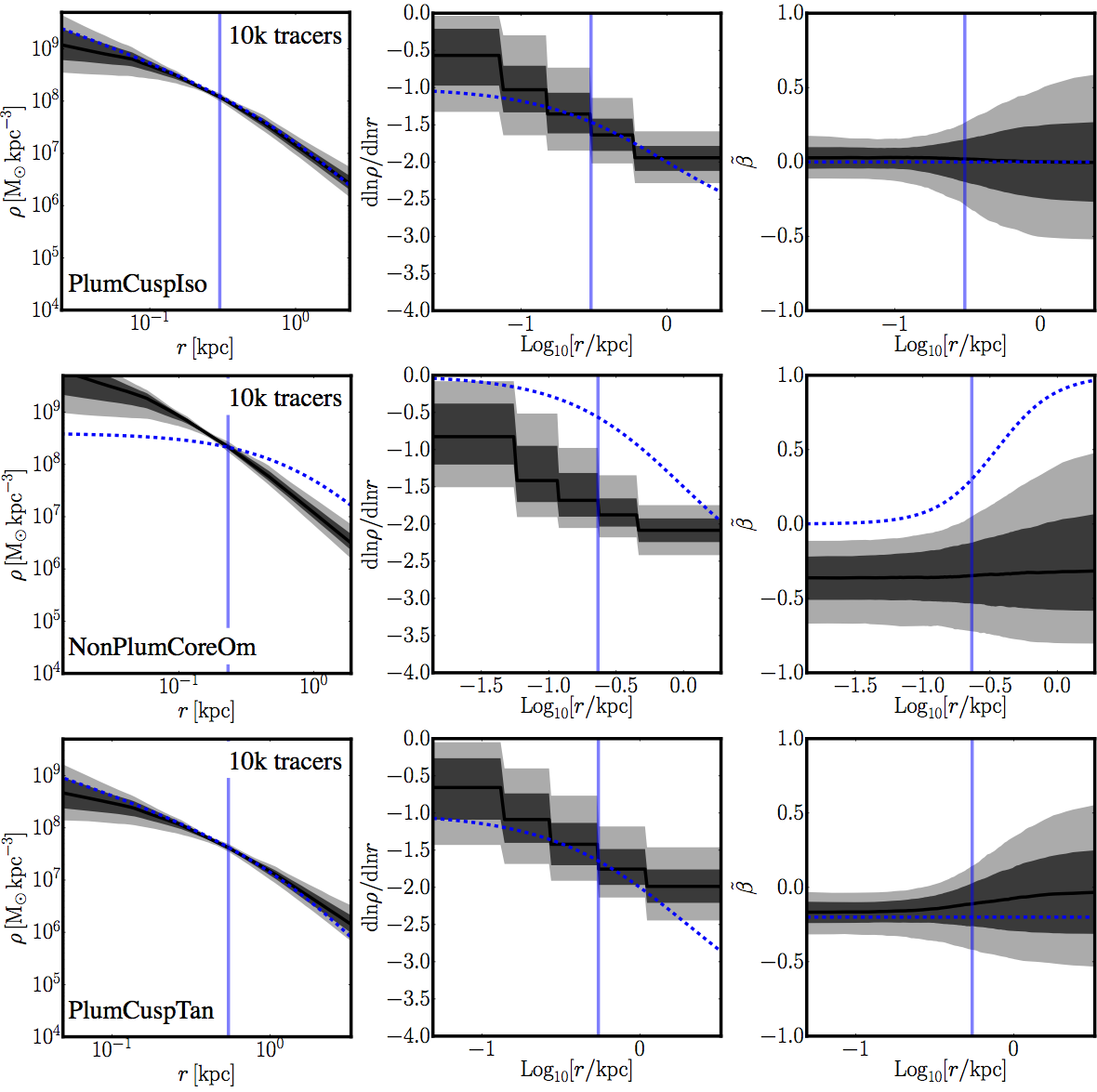}
\caption{As Figure \ref{fig:LOS_plots}, but using 10,000 tracer stars.}
\label{fig:10k_plots} 
\end{center}
\end{figure*}

\subsection{Line of sight velocities only}\label{sec:losonly}

We first consider how well \GravSphere\ performs when fitting only line-of-sight velocities, $\siglosobs(R)$, and the surface density of tracer stars $\Sigmaobs(R)$. The results are shown in Figures \ref{fig:FitExample} (top panels), \ref{fig:LOS_plots} (using 1000 tracer stars) and \ref{fig:10k_plots} (using 10,000 tracer stars).

In Figure \ref{fig:FitExample} (top panels), we show how well $\siglosobs(R)$ and $\Sigmaobs(R)$ are fit by \GravSphere\ for the NonPlumCoreOm mock with 1000 tracers (\GravSphere\ performs similarly well for the other mocks and so we omit these plots for brevity). In Figure \ref{fig:LOS_plots}, we show the radial density profile $\rho(r)$ (left), logarithmic density slope $d\ln\rho/d\ln r$ (middle) and symmetrised velocity anisotropy $\betastar$ (right) for the PlumCuspIso (top), NonPlumCoreOm (middle) and PlumCuspTan (bottom) mocks (see \S\ref{sec:mocks} and Table \ref{tab:mocks} for details of these mock data sets) that result from this fit. In Figure \ref{fig:10k_plots}, we show the same but for 10,000 tracers.

As can be seen, in all three mocks, we obtain a good estimate of the density at $R_{1/2}$, but a poor measure of both the radial profile of $\rho(r)$ and the velocity anisotropy profile $\betastar(r)$. This is true {\it even when using a rather optimistic 10,000 tracer stars} (Figure \ref{fig:10k_plots}), highlighting the familiar $\rho-\beta$ degeneracy (see \S\ref{sec:intro} and \citealt{2009ApJ...704.1274W,2010MNRAS.406.1220W}). In particular, the logarithmic derivative of the density, $d\ln\rho/d\ln r$ (middle panels), is poorly recovered, with only very weak constraints inside $r < R_{1/2}$ at 95\% confidence (light grey shaded regions). (Notice that this parameter moves in discrete steps. This owes to our parameterised form for $\rho_\DM$ that we described in detail in \S\ref{sec:Mmodel}.)

Finally, notice that our recovery for the NonPlumCoreOm mock (compare the shaded regions with the blue dashed lines in Figure \ref{fig:LOS_plots}, middle row) is biased, with the bias becoming even slightly worse as we move from 1000 to 10,000 tracers (Figure \ref{fig:10k_plots}, middle row). This is our worst mock from the whole default spherical-triaxial suite (\S\ref{sec:mocks}) which is why we have included it here. It is also the {\it only} mock from the full suite for which we become biased in this way. The problem occurs because this particular mock is both very cored (with $0 < \gamma(r) < 0.5$ for $r < R_{1/2}$) and very radially biased. Furthermore, the radial bias sets in only far beyond $R_{1/2}$ where there are few tracer stars (see Figure \ref{fig:LOS_plots}, middle row, middle and rightmost panels). In this case, the hypervolume of solution space remains large due to the $\rho-\beta$ degeneracy, but the correct solution lies right on the edge of this space. Correct models {\it are} found within our chains, but they are rare. This is not because these models are a poor representation of the data, but rather because there are many more models that are equally good representations of the data and that have $\betastar < 0$. This leads to the tangential bias on $\betastar$ that can be seen in Figures \ref{fig:LOS_plots} and \ref{fig:10k_plots}, with an associated bias on $\rho$ that it becomes too steep. 

The above presents us with a good opportunity to remind the reader that the shaded regions in Figure \ref{fig:LOS_plots} are {\it marginalised histograms} of the model parameters in radial bins. They do not represent any single model and, indeed, it can be the case that no model actually passes through the black median line (though in practice, this is not the case for the mock data we consider in this paper). These marginalised histograms are a convolution of both the data and our prior. Where the data constraints are poor, as is the case when only line-of-sight velocities are available, then the shaded regions that result depend as much on our choice of priors as the data. (A prior, for example, that $\beta_\infty = 1$ would remove the bias from the NonPlumCoreOm mock. But it would introduce a bias into the isotropic and tangential models.)

Ultimately, the only solution to the above bias problem is improved data. In the following subsections, we explore three methods that have been proposed in the literature to break the $\rho-\beta$ degeneracy by introducing additional data constraints: multiple tracer populations with distinct $R_{1/2}$ (\S\ref{sec:split}); VSPs (\S\ref{sec:vs}) and proper motions (\S\ref{sec:proper}).

\subsection{Split populations}\label{sec:split}

\begin{figure*}
\begin{center}
\includegraphics[width=0.75\textwidth]{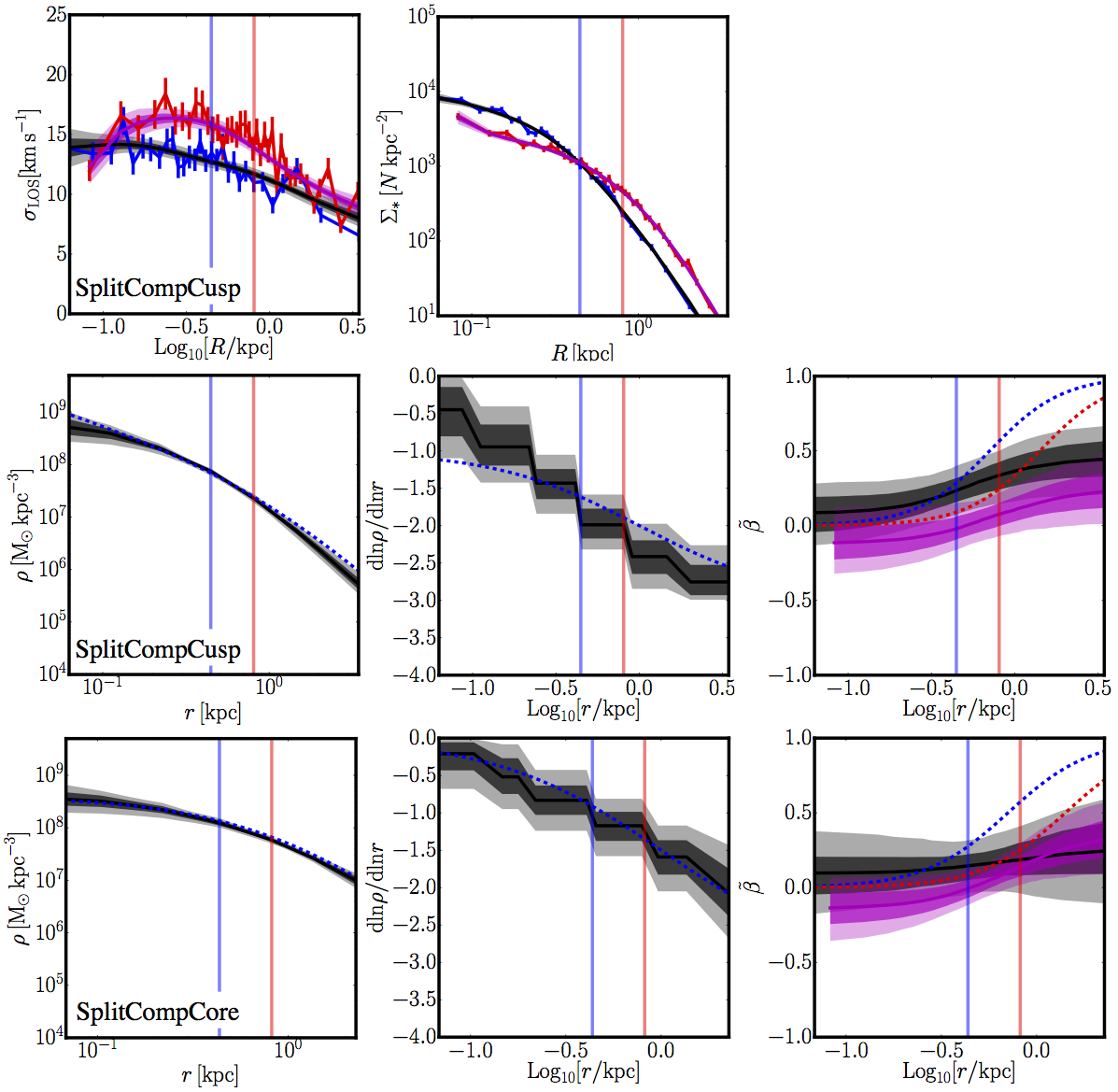}
\caption{Testing \GravSphere\ with split population data. {\bf Top row:} example fits for the SplitCompCusp mock (see Table \ref{tab:mocks}), showing $\siglos(R)$ (left) and $\Sigma_*(R)$ (middle) for the inner (blue) and outer (red) tracer populations. The lines and shaded regions are as in Figure \ref{fig:FitExample}, except that now there are two vertical lines to mark the $R_{1/2}$ of the inner (blue) and outer (red) population, while the grey and magenta shaded bands show the 68\% and 95\% confidence intervals of our fits for the inner and outer population, respectively. {\bf Middle \& bottom rows:} the radial density profile $\rho(r)$; logarithmic slope of the density profile $d\ln\rho/d\ln r$; and symmetrised velocity anisotropy parameter $\betastar$ for the SplitCompCusp mock (middle) and SplitCompCore mock (bottom). The lines and shaded regions are as in Figure \ref{fig:LOS_plots}, but with the inner (vertical blue) and outer (vertical red) $R_{1/2}$ of the two populations marked. In the rightmost panels, we show results for the inner $\betastar_1$ (grey shaded region) and outer $\betastar_2$ (magenta shaded region); the correct solutions are marked by the blue and red dashed lines, respectively. See \S\ref{sec:multimethod} and \S\ref{sec:split} for further details.}
\label{fig:Split_plots} 
\end{center}
\end{figure*}

As discussed in \S\ref{sec:intro}, a popular method for breaking $\rho-\beta$ degeneracy is to simultaneously model multiple tracers with distinct $R_{1/2}$ \citep[e.g.][]{2000AJ....119..153S,2008ApJ...681L..13B,2011ApJ...742...20W,2011MNRAS.tmp.1606A,2014MNRAS.439..659N}. Each tracer $i$ provides a robust measurement of $\rho(R_{1/2,i})$ (c.f. \S\ref{sec:losonly}), allowing the density profile to be pinned down at multiple radii.

While the idea of using split populations to break the $\rho-\beta$ degeneracy is clear, there has been some debate in the literature over its efficacy (e.g. \citealt{2008ApJ...681L..13B,2011ApJ...742...20W,2013MNRAS.433.3173B,2014ApJ...791L...3B} and see the discussion in \S\ref{sec:intro}). In this context, \GravSphere\ is particularly useful since it assumes only that the system is close to spherical, that the tracers are in Virial equilibrium, and that the tracers have been perfectly separated (see \S\ref{sec:method}).

In Figure \ref{fig:Split_plots}, we show results for our two split population mocks: SplitCompCusp and SplitCompCore (see Table \ref{tab:mocks}). The top row shows an example of the data fit for the SplitCompCusp case (the SplitCompCore mock is fit similarly well and so we have omitted these plots for brevity). The lines and shaded regions are as in Figure \ref{fig:FitExample}, except that now there are two vertical lines to mark the $R_{1/2}$ of the inner (blue) and outer (red) population, while the grey and magenta shaded bands show the 68\% and 95\% confidence intervals of our fits for the inner and outer population, respectively. The middle and bottom rows show the radial density profile $\rho(r)$; logarithmic slope of the density profile $d\ln\rho/d\ln r$; and symmetrised velocity anisotropy parameter $\betastar$ for the SplitCompCusp mock (middle) and SplitCompCore mock (bottom). The lines and shaded regions are as in Figure \ref{fig:LOS_plots}, but with the inner (vertical blue) and outer (vertical red) $R_{1/2}$ of the two populations marked. In the rightmost panels, we show results for the inner $\betastar_1$ (grey shaded region) and outer $\betastar_2$ (magenta shaded region); the correct solutions are marked by the blue and red dashed lines, respectively.

As can be seen from Figure \ref{fig:Split_plots}, using two populations performs as advertised, dramatically improving the recovery\footnote{Since the total number of tracers in these split mocks is 6,580 for SplitCompCore and 7,105 for SplitCompCusp (see Table \ref{tab:mocks} and \S\ref{sec:mocks}), we should compare the performance of \GravSphere\ on these with the 10k single component models in Figure \ref{fig:10k_plots}.} of $\rho(r)$. We obtain an excellent recovery of $\rho(r)$ in-between $R_{1/2,1}$ and $R_{1/2,2}$, with tight constraints also over the range $0.5 R_{1/2,1} < r < 2 R_{1/2,2}$. Inside $r < 0.5 R_{1/2,1}$ the constraints become poorer again and we are not able to well-measure the very inner logarithmic cusp slope. (Note, however, that this does depend on our choice of mass model and prior. If we assume a single power law inside $R_{1/2,1}$ then this will be very well constrained, as pointed out by \citealt{2011ApJ...742...20W}. Similarly, our constraints are improved by our monotonicity prior on $d\ln \rho_{\rm dm} / d\ln r$ (\S\ref{sec:Mmodel}) since this forbids an inner density slope that is steeper than the outer density slope.)

Interestingly, however, while the split component data provide an excellent recovery of $\rho(r)$, $\betastar_1$ and $\betastar_2$ are poorly recovered (see Figure \ref{fig:Split_plots}, middle and bottom rows, rightmost panels). The inner population is reasonably well recovered inside $r < R_{1/2,1}$ within our 95\% confidence intervals, but for $r > R_{1/2,1}$ we favour only weakly radially anisotropic models. Similarly, the outer population is poorly recovered, with a weak tangential bias on $\betastar_2$ for $r < R_{1/2,2}$ and only a hint of radial anisotropy detected at larger radii. It is interesting that despite this $\rho(r)$ is not biased, suggesting that it benefits from the torsioning of one population against the other. We defer a discussion of how \GravSphere\ performs when populations are imperfectly split to future work. 

\subsection{Virial Shape Parameters}\label{sec:vs}

\begin{figure*}
\begin{center}
\includegraphics[width=0.75\textwidth]{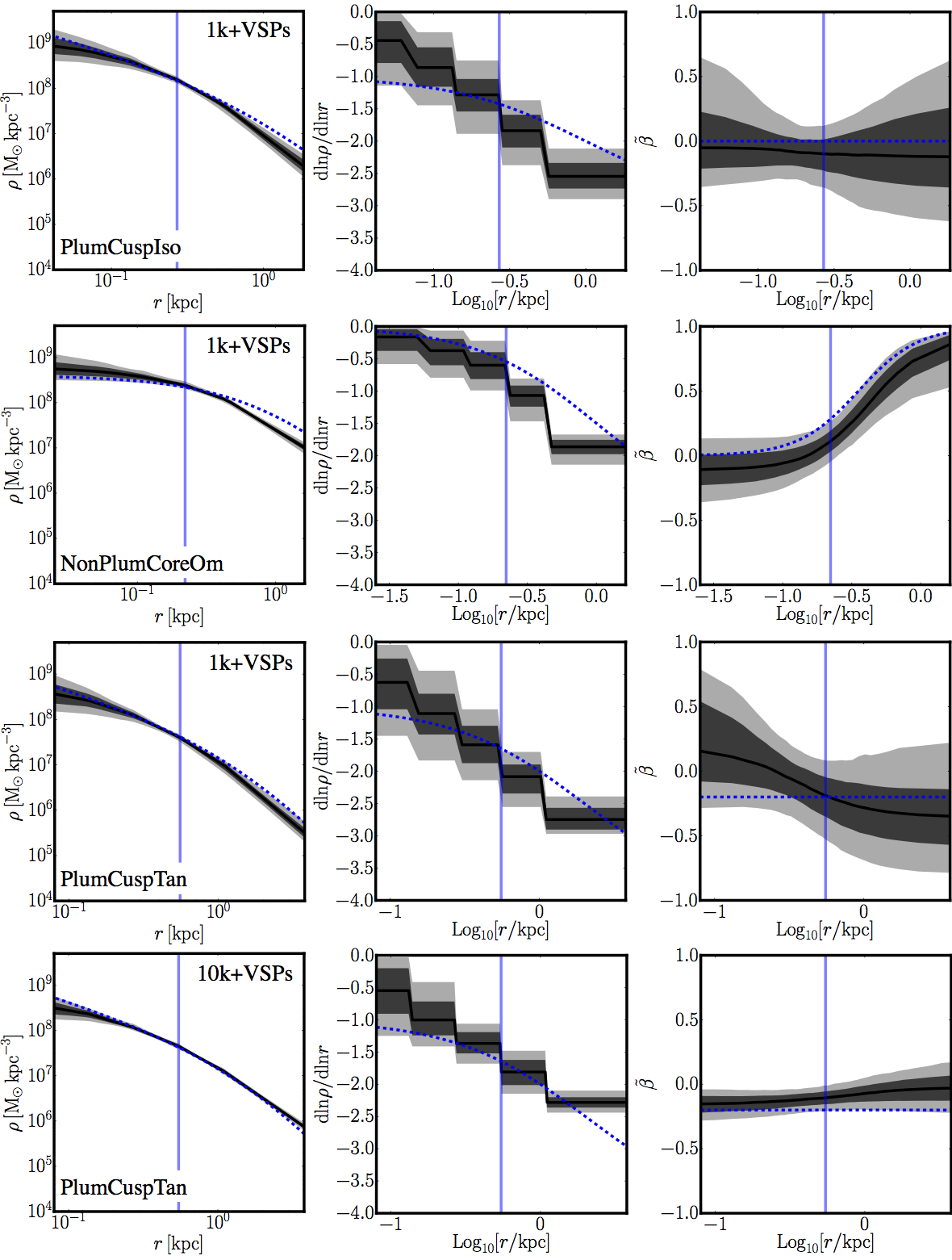}
\caption{As Figure \ref{fig:LOS_plots}, but including constraints from the `Virial Shape Parameters' (VSPs; see \S\ref{sec:VSPsmethod} and \S\ref{sec:vs}). The top three rows show results using 1000 tracer stars; the bottom row shows what happens when we use 10k tracer stars for the PlumCuspTan mock. An example model fit to data for these is shown in Figure \ref{fig:FitExample} (middle row) for the NonPlumCoreOm mock.}
\label{fig:vs_plots} 
\end{center}
\end{figure*}

Split populations appear to be a promising avenue for breaking the $\rho-\beta$ degeneracy. However, unambiguously split populations are not always readily available. Furthermore, as we showed in \S\ref{sec:split}, they do not yield strong constraints on $\beta$. In this subsection, we consider instead the utility of the fourth order `Virial Shape Parameters' (VSPs) that we introduced in \S\ref{sec:VSPsmethod}, $\vsone$ and $\vstwo$ (see equations \ref{eqn:vs1} and \ref{eqn:vs2}). 

In Figure \ref{fig:vs_plots}, we show results for the single component mocks PlumCuspIso, NonPlumCoreOm and PlumCuspTan, but now including constraints from $\vsone$ and $\vstwo$. The top three rows show results using 1000 tracer stars; the bottom row shows what happens when we use 10k tracer stars for the PlumCuspTan mock. The lines and shaded regions are as in Figure \ref{fig:LOS_plots}. We show an example fit for the NonPlumCoreOm mock in Figure \ref{fig:FitExample} (middle row), where the middle and right panels show marginalised histograms of $\vsone$ and $\vstwo$, respectively. The two additional VSP data points that we fit are marked on these histograms, with errors calculated as in \S\ref{sec:mcmcmethod}. (For brevity, we omit the similar plots for the PlumCuspIso and PlumCuspTan mocks since \GravSphere\ performs similarly well on these.)

As can be seen from Figure \ref{fig:vs_plots}, even with just 1000 tracers, including the VSPs produces a dramatic improvement in the performance of \GravSphere. The density $\rho(r)$ is well recovered over the range $0.5 R_{1/2} < r < 2 R_{1/2}$, with poorer constraints for $r < 0.5 R_{1/2}$ and a small bias creeping in at larger radii. This bias is not a particular cause for concern since there are few tracers this far out. For the PlumCuspIso and PlumCuspTan mocks with 1000 tracers, the $\rho-\beta$ degeneracy is only weakly broken. However, for the NonPlumCoreOm mock, the improvement is dramatic. The VSPs are able to detect this mock's strong radial anisotropy, reproducing the inner isotropic $\beta(0) \sim 0$ and its sharp radially anisotropic rise to large $r$ (see Figure \ref{fig:vs_plots}, middle row, rightmost panel). 

With 10,000 tracers, the VSPs perform even better. Figure \ref{fig:vs_plots}, bottom row, shows results for the PlumCuspTan mock using 10k tracers (similar results are obtained with 10k tracers for the other mocks and so we omit these for brevity). We now obtain tight constraints on $\rho$ over the range $0.25 R_{1/2} < r < 4R_{1/2}$ at better than 95\% confidence. In particular, this leads to an excellent recovery of $d\ln\rho/d\ln r(r)$ (middle panel, bottom row of Figure \ref{fig:vs_plots}) and $\betastar(r)$ (rightmost panel, bottom row of Figure \ref{fig:vs_plots}).

Finally, we note that combining VSPs with split populations may yield even tighter constraints. We will explore this idea further in future work.

\subsection{Including proper motion data}\label{sec:proper}

\begin{figure*}
\begin{center}
\includegraphics[width=0.75\textwidth]{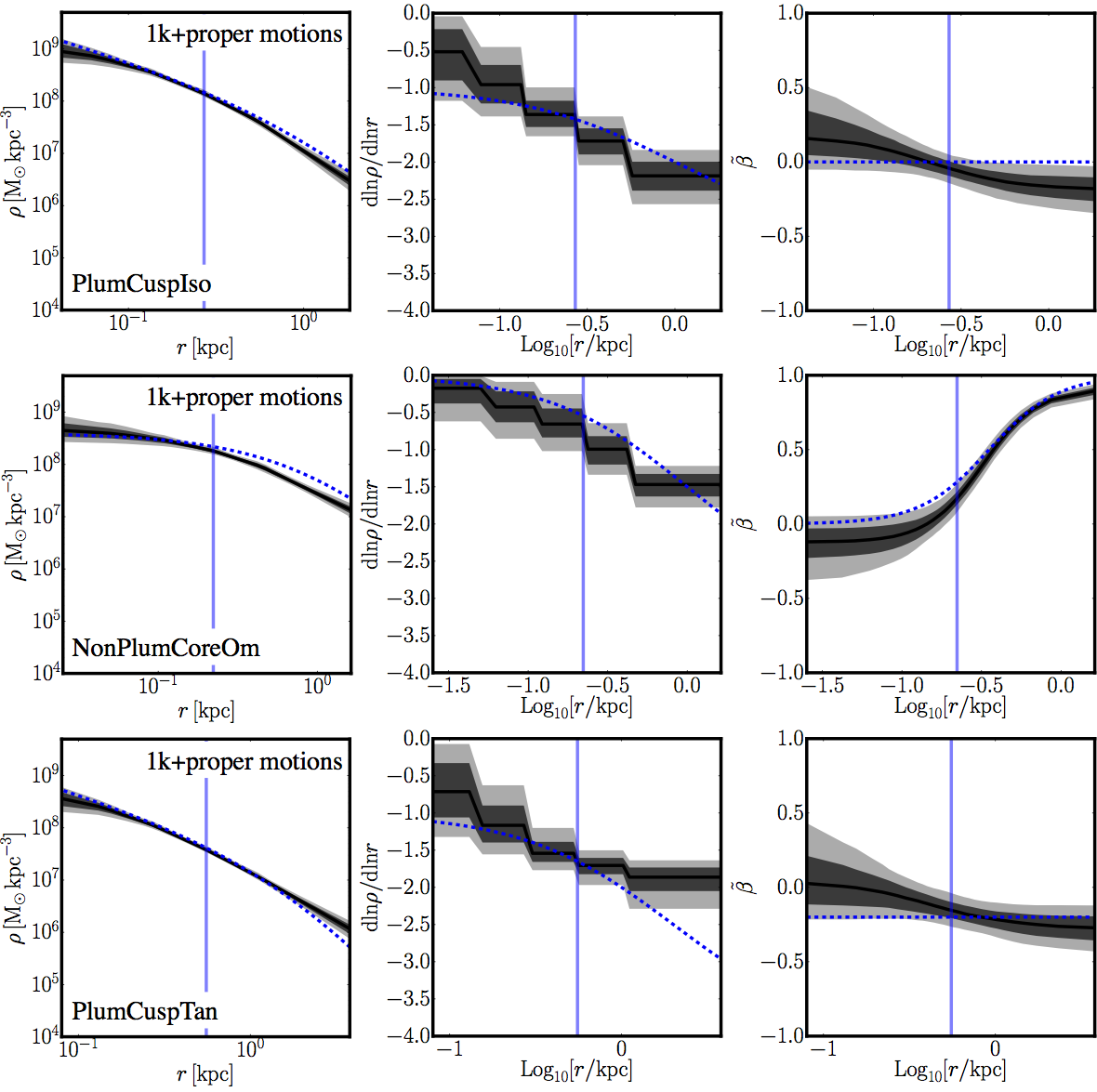}
\caption{As Figure \ref{fig:LOS_plots}, but including constraints from the tangential and radial velocity dispersions, $\sigpmt$ and $\sigpmr$, that are derived from stellar proper motions (see \S\ref{sec:propermethod} and \S\ref{sec:proper}). An example model fit to these data is shown in Figure \ref{fig:FitExample} (bottom row) for the NonPlumCoreOm mock.}
\label{fig:Prop_plots} 
\end{center}
\end{figure*}

In this subsection, we consider the utility of adding proper motion data. In projection, this means adding the tangential ($\sigpmt$) and radial ($\sigpmr$) velocity dispersion constraints, as described in \S\ref{sec:propermethod}. Our results for the PlumCuspIso, NonPlumCoreOm and PlumCuspTan mocks are shown in Figure \ref{fig:Prop_plots}. The lines and shaded regions are as in Figure \ref{fig:LOS_plots}. An example model fit to these data is shown in Figure \ref{fig:FitExample} (bottom row) for the NonPlumCoreOm mock (\GravSphere\ performed similarly well on the other mocks and so we omit their similar plots for brevity.)

As can be seen from Figure \ref{fig:Prop_plots}, even with 1000 tracers stars \GravSphere\ gives an excellent recovery of $\rho(r)$, $d\ln\rho/d\ln r$ and $\betastar$, once proper motion data are added. As discussed in \S\ref{sec:propermethod}, the addition of $\sigpmt$ and $\sigpmr$ should provide a strong breaking of the $\rho-\beta$ degeneracy and this is reflected in the performance of \GravSphere\ in Figure \ref{fig:Prop_plots}. As was the case when adding VSPs (Figure \ref{fig:vs_plots}), the NonPlumCoreOm mock becomes slightly biased at large radii. Otherwise, the results for all three mocks are excellent, recovering the input models within our 95\% confidence intervals. 

While proper motion data clearly gives the best performance, collecting data of the quality assumed in Figure \ref{fig:Prop_plots} will require a new and dedicated satellite astrometric mission \citep[e.g.][]{2007ApJ...657L...1S}. Even then, we would only be able to study to the very nearest stellar systems. By contrast, both split populations and VSPs can be used on existing data and will continue to improve as spectra for more and more tracers are collected.

\subsection{Triaxial Mock Data}\label{sec:triaxial}
\begin{figure*}
\begin{center}
\includegraphics[width=0.75\textwidth]{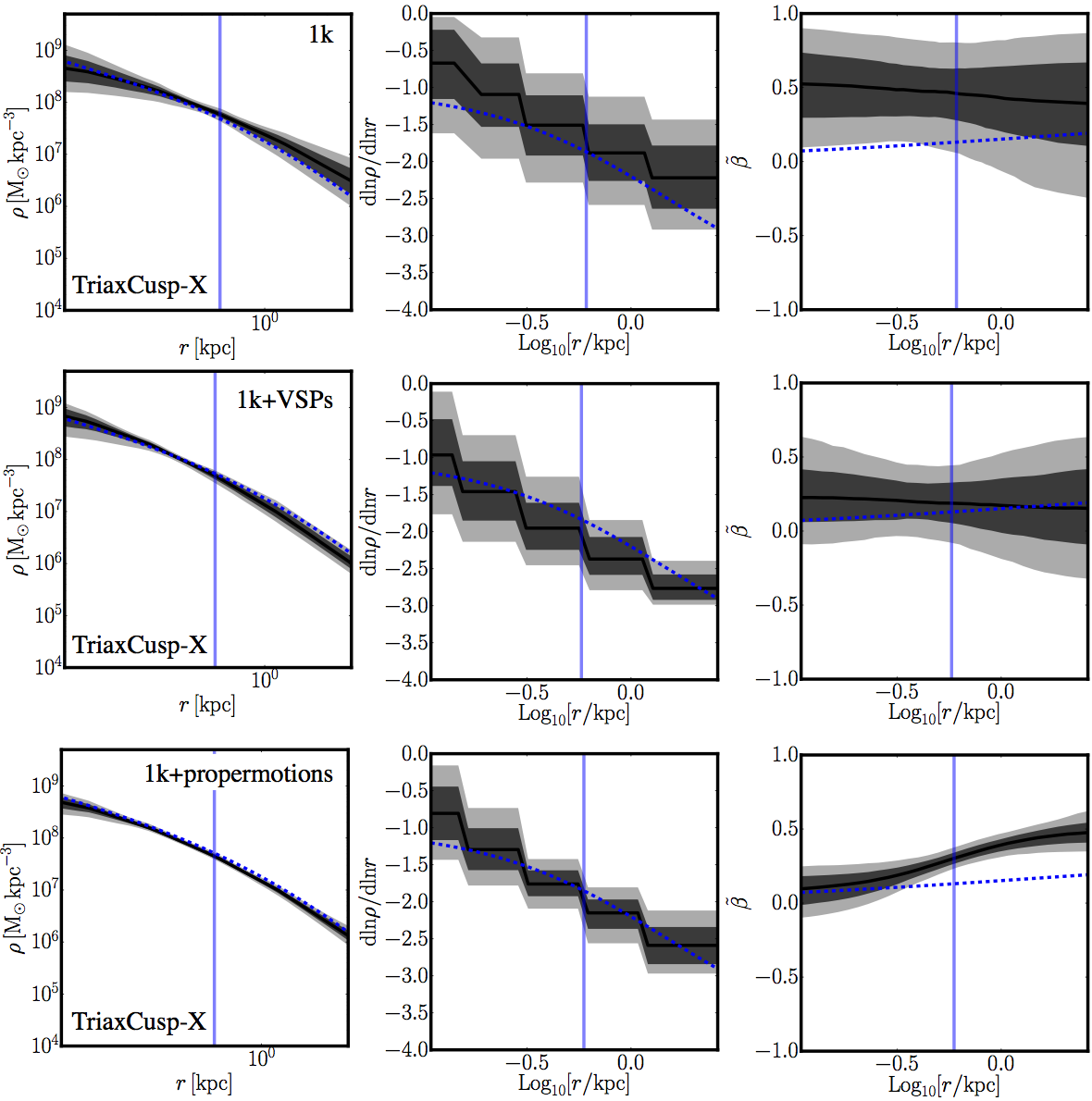}
\caption{Testing \GravSphere\ with triaxial mock data looking down the long axis ($X$). From top to bottom, the panels show results for the TriaxCusp mock (Table \ref{tab:mocks}) for 1000 tracers with line-of-sight velocities only; including VSPs; and including proper motions. The lines and shaded regions are as in Figure \ref{fig:LOS_plots}.}
\label{fig:Triax_cuspX} 
\end{center}
\end{figure*}

\begin{figure*}
\begin{center}
\includegraphics[width=0.75\textwidth]{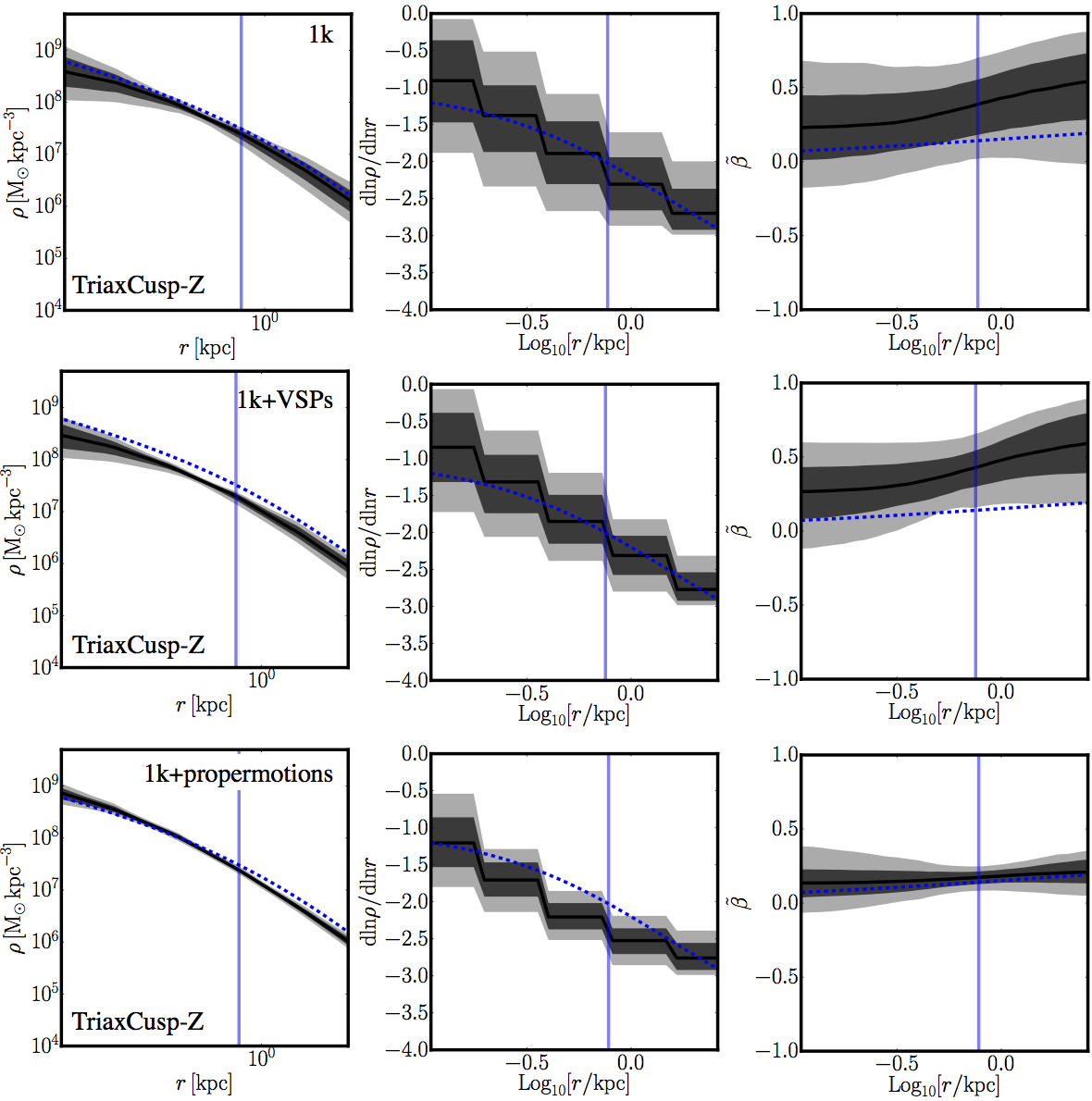}
\caption{As Figure \ref{fig:Triax_cuspX}, but looking down the short axis ($Z$).}
\label{fig:Triax_cuspZ} 
\end{center}
\end{figure*}

\begin{figure*}
\begin{center}
\includegraphics[width=0.75\textwidth]{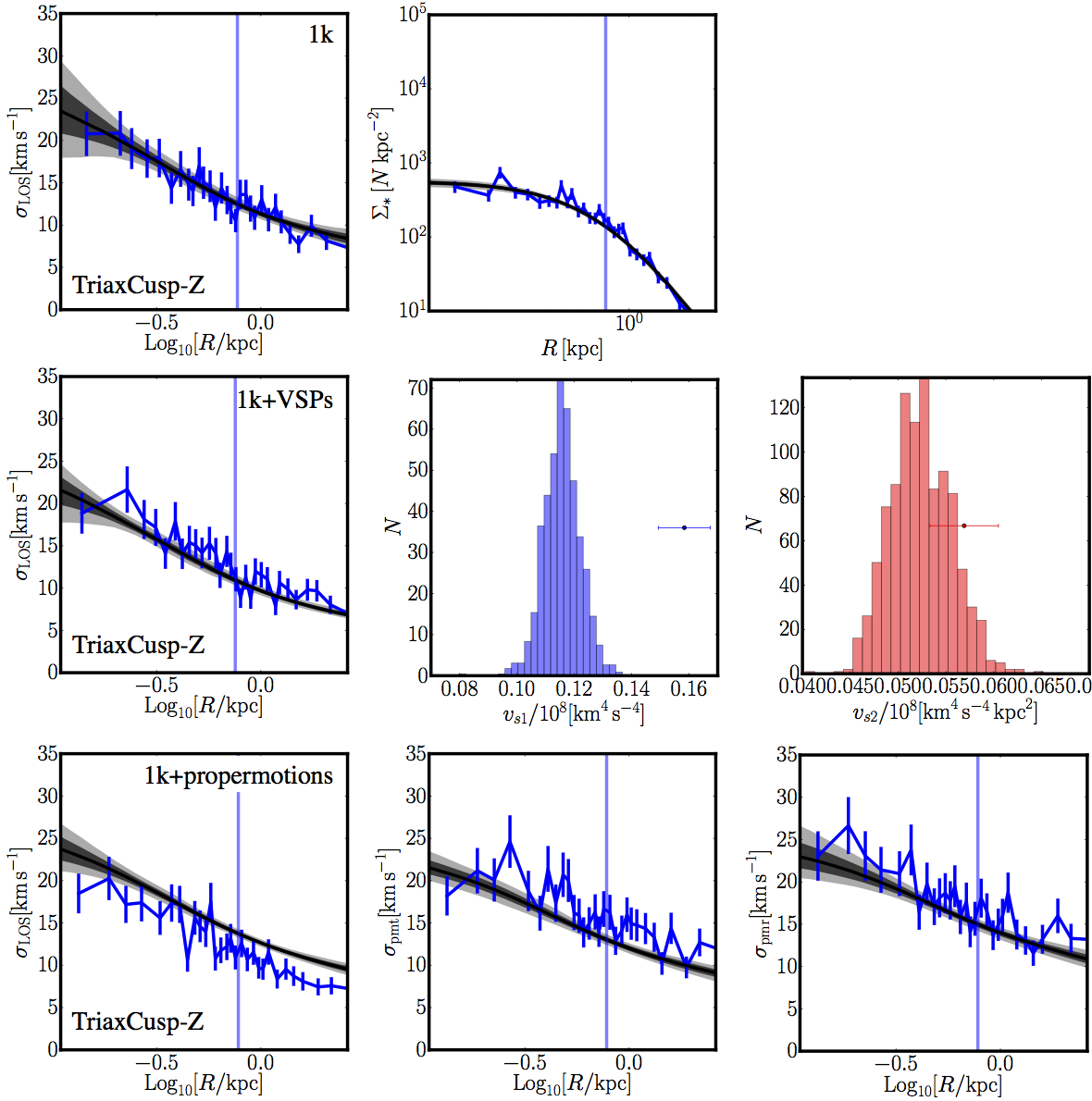}
\caption{\GravSphere\ model fits for the TriaxCusp mock data looking down the short axis ($Z$). The panels, lines and shaded regions are as in Figure \ref{fig:FitExample}. The rows show model fits for, from top to bottom, 1000 tracers with line-of-sight velocities only; including VSPs; and including proper motions. Notice that the fits are visibly poor for the case where VSP or proper motion data are included (middle and bottom row). This suggests that there is power in the data to determine that the system is not spherical.}
\label{fig:Triax_cuspZ_fit} 
\end{center}
\end{figure*}

Finally, we consider in this subsection how \GravSphere\ performs on {\it triaxial} mock data for which we expect it to become biased (because the method fundamentally assumes spherical symmetry; see \S\ref{sec:method}). In Figure \ref{fig:Triax_cuspX}, we show results for the TriaxCusp mock (see Table \ref{tab:mocks}) looking down the long axis ($X$) and in Figure \ref{fig:Triax_cuspZ}, we show the same looking down the short axis ($Z$). The lines and shaded regions are as in Figure \ref{fig:LOS_plots}. From top to bottom, the panels show results for 1000 tracers with line-of-sight velocities only; including VSPs; and including proper motions. We show the model fits to the data for the short axis case in Figure \ref{fig:Triax_cuspZ_fit} (the long axis behaves similarly).

Firstly, notice that when staring down the long axis, $\rho(r)$ is recovered very well with no noticeable bias (compare the grey shaded regions and blue dashed lines in the left and middle columns of Figure \ref{fig:Triax_cuspX}). Only when adding full proper motion data do the models become biased at greater than 95\% confidence, and then only in $\betastar$ (see the bottom right panel of Figure \ref{fig:Triax_cuspX}). 

The performance of \GravSphere\ is slightly worse when looking along the short axis, however (Figure \ref{fig:Triax_cuspZ}). Now when adding VSPs. the density becomes biased low at all radii, while a similar albeit weaker effect is also seen when adding the proper motion data. Interestingly, however, in both cases we would be able to detect that something has gone wrong. As can be seen in Figure \ref{fig:Triax_cuspZ_fit}, the actual fits to the data are excellent when only line-of-sight velocities are available (top row). However, when adding either VSPs (middle row) or proper motions (bottom row), the fits become visibly poor. In the case of VSPs, the parameter $\vsone$ can no longer be fit, while for the proper motions, $\siglos$ becomes visibly biased high as compared to the data. A corollary of this is that there appears to be power in the data to determine that the system is not spherical. We will explore this further in future work.

Finally, notice that although \GravSphere\ becomes slightly biased when applied to triaxial mock data, in almost all cases the density at $R_{1/2}$ remains well-recovered. This is consistent with recent work by \citet{2016ApJ...830L..26S} who show that Jeans estimators remain valid for stellar systems that are flattened along constant ellipsoids.

\section{Conclusions}\label{sec:conclusions}
We have presented a new non-parametric Jeans method -- \GravSphere\ -- and tested it on a wide range of mock data. Our goal was to assess how well three popular methods for breaking the $\rho-\beta$ degeneracy work: split populations; `Virial Shape Parameters' (VSPs); and proper motions. Our key results are as follows:

\begin{itemize} 

\item We confirm that with only line-of-sight velocity data, \GravSphere\ provides a good estimate of the density at the projected stellar half mass radius, $\rho(R_{1/2})$, but is not able to measure $\rho(r)$ or $\beta(r)$, even with 10,000 tracer stars.

\item We find that two populations provide an excellent recovery of $\rho(r)$ in-between their respective $R_{1/2}$. However, even with a total of $\sim 7,000$ tracers, we are not able to well-constrain $\beta(r)$ for either population. 

\item Using 1000 tracers with higher order VSPs, we are able to measure $\rho(r)$ over the range $0.5 < r/R_{1/2} < 2$ and broadly constrain $\beta(r)$. With 10,000 tracers, we obtain an exquisite measure of $\rho$ and $\beta$ over the range $0.25 R_{1/2} < r < 4R_{1/2}$ at better than 95\% confidence. 

\item Including proper motion data for all stars gives an even better performance than using VSPs. With just 1000 tracers, we can measure $\rho$ and $\beta$ over the range $0.25 < r/R_{1/2} < 4$. However, obtaining these data will require a specialised astrometric mission. For the foreseeable future, this will only be possible for the very nearest stellar systems.

\item We tested \GravSphere\ on a triaxial mock galaxy that has axis ratios typical of a merger remnant, $[1:0.8:0.6]$. In this case, with 1000 tracers and using VSPs or proper motions, \GravSphere\ can become slightly biased. However, we find that when this occurs the data are poorly fit, allowing us to detect when such departures from spherical symmetry become problematic. A corollary of this is that there appears to be power in the data to determine that the system is not spherical. We will explore this further in future work.

\end{itemize} 

\section{Acknowledgments}\label{sec:acknowledgements}
JIR would like to acknowledge support from SNF grant PP00P2\_128540/1; STFC consolidated grant ST/M000990/1; and the MERAC foundation. We would like to thank the \GC\ initiative for many stimulating workshops that led to the mock data compilation that we have used here. In particular, JIR would like to thank the Nordita institute for hosting the last \GC\ meeting at which these ideas came to fruition. We would like to thank Matt Walker, Jorge Pe\~narrubia, Mark Gieles, Michelle Collins, Payel Das, Laura Watkins, Gary Mamon, Eugene Vasiliev and John Magorrian for useful feedback and stimulating discussions.
\appendix

\section{The choice of binning}\label{app:binning}

\begin{figure*}
\begin{center}
\includegraphics[width=0.75\textwidth]{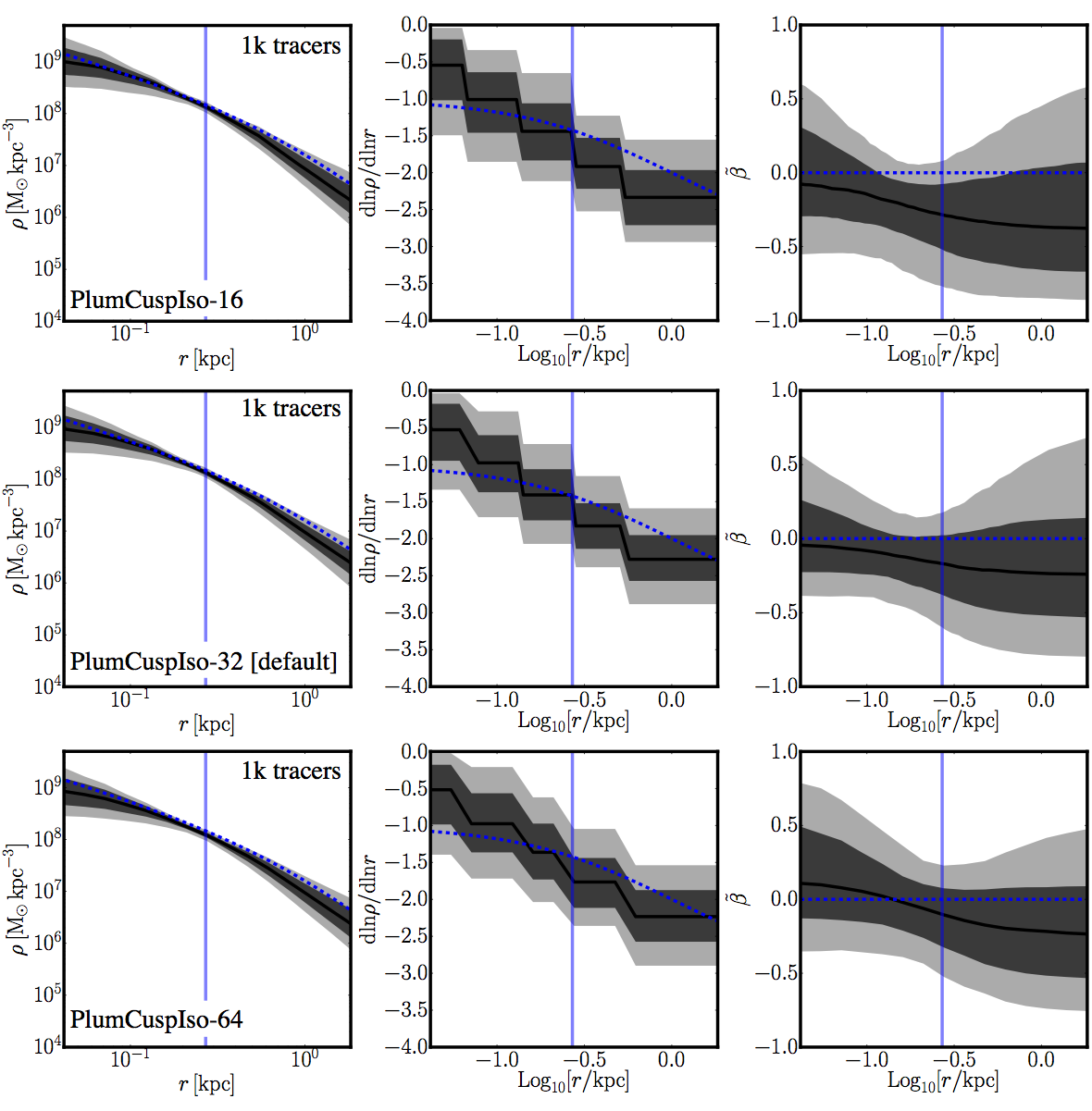}
\caption{Testing the effect of our choice of binning on the results. From top to bottom, the rows show results for the PlumCuspIso mock using 1000 tracer stars with only line-of-sight velocity data. The lines and shaded regions are as in Figure \ref{fig:LOS_plots}. The top row shows the results using $N_{\rm bin} = 16$ stars per bin; the middle for our default $N_{\rm bin} = \sqrt{N_*} = 32$ stars per bin; and the bottom for $N_{\rm bin} = 64$ stars per bin. As can be seen, these changes in binning produce no noticeable effect on the results.}
\label{fig:Bintest_plots} 
\end{center}
\end{figure*}

In this Appendix, we explore the sensitivity of our results to our choice of data binning. By default, we use $N_{\rm bin} = \sqrt{N_*}$ bins where $N_*$ is the total number of stellar tracers in the data (see \S\ref{sec:mcmcmethod}). For 1000 tracers, this gives $N_{\rm bin} = 32$. In Figure \ref{fig:Bintest_plots}, we show how our results for the PlumCuspIso mock change when we use instead $N_{\rm bin} = 16$ (top row) and $N_{\rm bin} = 64$ (bottom row), assuming only line-of-sight velocities for 1000 tracer stars. As can be seen, these changes in binning produce no noticeable effect on the results.

\section{Virial Shape Parameter Estimators}\label{app:estimators}

In this work, we estimate $\vsone$ and $\vstwo$ from the data by numerically integrating the observed $\vlosfour$ over our model-fitted $\Sigma_*$ (see equations \ref{eqn:vs1data} and \ref{eqn:vs2data}). We then estimate the errors on $\vsone$ and $\vstwo$ using a Monte-Carlo sampling method described in \ref{sec:mcmcmethod}. In this appendix, we compare this procedure with the estimators given in \citet{2014MNRAS.441.1584R}:
\begin{eqnarray}
\vsone = \int_0^{\infty} \Sigma_* \vlosfour R dR & = & \frac{1}{2\pi}\int_0^{\infty} \vlosfour dN  \nonumber \\
& \simeq & \frac{1}{2\pi}\sum_i^{N_s} v_{{\rm LOS},i}^4
\label{eqn:RFvsone}
\end{eqnarray}
where we recall that $\Sigma_*$ is the surface number density of stars and $dN = 2\pi \Sigma_* R dR$. Similarly:

\begin{equation} 
\vstwo = \frac{1}{2\pi} \sum_i^{N_s} v_{{\rm LOS},i}^4 R_i^2
\label{eqn:RFvstwo}
\end{equation}
We estimate errors on the \citet{2014MNRAS.441.1584R} estimators by Monte-Carlo sampling over the errors on the individual $v_{{\rm LOS},i}$ measurements and calculating the variance in the solutions to equations \ref{eqn:RFvsone} and \ref{eqn:RFvstwo}.

\begin{figure}
\begin{center}
\includegraphics[width=0.45\textwidth]{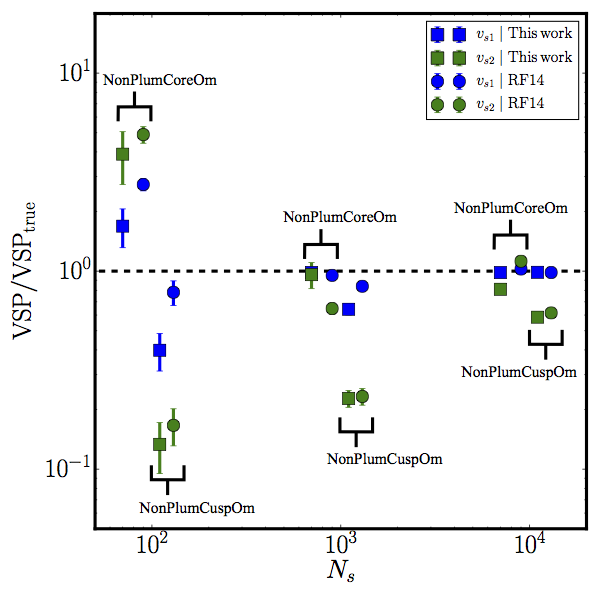}
\vspace{-2mm}
\caption{A comparison of the VSP estimators in this paper (squares) with those from \citet{2014MNRAS.441.1584R} (circles; equations \ref{eqn:RFvsone} and \ref{eqn:RFvstwo}). We plot the estimated over the true VSP (as calculated from the mock density, tracer density and velocity anisotropy profiles), VSP/VSP$_{\rm true}$, as a function of the number of tracer stars, $N_s = 100, 1000$ and $10,000$. The data points are slightly offset from one another horizontally to aid clarity. The horizontal dashed line marks VSP/VSP$_{\rm true} = 1$, corresponding to zero bias. The blue data points show results for $\vsone$, the green for $\vstwo$. We show results for two different \GC\ mock data sets, NonPlumCoreOm and NonPlumCuspOm, as marked (see Table \ref{tab:mocks}).}
\label{fig:vsp_bias} 
\end{center}
\end{figure}

In Figure \ref{fig:vsp_bias}, we compare our estimators for $\vsone$ and $\vstwo$ with those in equations \ref{eqn:RFvsone} and \ref{eqn:RFvstwo}. We plot the estimated over the true VSP (as calculated from the mock density, tracer density and velocity anisotropy profiles), VSP/VSP$_{\rm true}$, as a function of the number of tracer stars, $N_s = 100, 1000$ and $10,000$. The data points are slightly offset from one another horizontally to aid clarity. The horizontal dashed line marks VSP/VSP$_{\rm true} = 1$, corresponding to zero bias. The blue data points show results for $\vsone$, the green for $\vstwo$. The squares show results from the estimators presented in this work; the circles show those from \citet{2014MNRAS.441.1584R} (equations \ref{eqn:RFvsone} and \ref{eqn:RFvstwo}). We show results for two different \GC\ mock data sets: the NonPlumCoreOm mock discussed already in this paper, and a similar mock but with a dark matter cusp, NonPlumCuspOm, as marked on the Figure (see Table \ref{tab:mocks}).

As can be seen in Figure \ref{fig:vsp_bias}, increasing the number of tracer stars, $N_s$, reduces the bias for both estimators. As might be expected, there is less bias for $\vsone$ (blue) than $\vstwo$ (green) since $\vstwo$ is much more sensitive to the fall-off of the light profile at large radii where there is little data. Our new estimators for $\vsone$ and $\vstwo$ perform similarly well to those from \citet{2014MNRAS.441.1584R}. As anticipated in \S\ref{sec:VSPsmethod}, the main advantage of our new estimators is a more realistic error estimate at low $N_s$. Finally, notice that the NonPlumCoreOm mock is less biased, particularly for $\vstwo$ than the NonPlumCuspOm mock. This is because the NonPlumCuspOm mock has the same dark matter scale length as NonPlumCoreOm but the stars are much more concentrated, with a scale length of $r_* = 0.1$\,kpc as compared to $r_* = 0.25$\,kpc. This means that at fixed $N_s$, there is correspondingly less information about the large radius behaviour of the mass distribution that $\vstwo$ is particularly sensitive to. We will discuss this mock in more detail in a companion paper where we describe the full \GC\ mock data suite (Read et al. in prep.).

\bibliographystyle{mnras}
\bibliography{refs,ref2}

\begin{thebibliography}{}
\makeatletter
\relax
\def\mn@urlcharsother{\let\do\@makeother \do\$\do\&\do\#\do\^\do\_\do\%\do\~}
\def\mn@doi{\begingroup\mn@urlcharsother \@ifnextchar [ {\mn@doi@}
  {\mn@doi@[]}}
\def\mn@doi@[#1]#2{\def\@tempa{#1}\ifx\@tempa\@empty \href
  {http://dx.doi.org/#2} {doi:#2}\else \href {http://dx.doi.org/#2} {#1}\fi
  \endgroup}
\def\mn@eprint#1#2{\mn@eprint@#1:#2::\@nil}
\def\mn@eprint@arXiv#1{\href {http://arxiv.org/abs/#1} {{\tt arXiv:#1}}}
\def\mn@eprint@dblp#1{\href {http://dblp.uni-trier.de/rec/bibtex/#1.xml}
  {dblp:#1}}
\def\mn@eprint@#1:#2:#3:#4\@nil{\def\@tempa {#1}\def\@tempb {#2}\def\@tempc
  {#3}\ifx \@tempc \@empty \let \@tempc \@tempb \let \@tempb \@tempa \fi \ifx
  \@tempb \@empty \def\@tempb {arXiv}\fi \@ifundefined
  {mn@eprint@\@tempb}{\@tempb:\@tempc}{\expandafter \expandafter \csname
  mn@eprint@\@tempb\endcsname \expandafter{\@tempc}}}

\bibitem[\protect\citeauthoryear{{Agnello} \& {Evans}}{{Agnello} \&
  {Evans}}{2012}]{2012ApJ...754L..39A}
{Agnello} A.,  {Evans} N.~W.,  2012, \mn@doi [\apjl]
  {10.1088/2041-8205/754/2/L39}, \href
  {http://adsabs.harvard.edu/abs/2012ApJ...754L..39A} {754, L39}

\bibitem[\protect\citeauthoryear{{Amorisco} \& {Evans}}{{Amorisco} \&
  {Evans}}{2011}]{2011MNRAS.tmp.1606A}
{Amorisco} N.~C.,  {Evans} N.~W.,  2011, \mn@doi [\mnras]
  {10.1111/j.1365-2966.2011.19684.x}, \href
  {http://adsabs.harvard.edu/abs/2011MNRAS.tmp.1606A} {p.~1606}

\bibitem[\protect\citeauthoryear{{An} \& {Evans}}{{An} \&
  {Evans}}{2006}]{2006ApJ...642..752A}
{An} J.~H.,  {Evans} N.~W.,  2006, \mn@doi [\apj] {10.1086/501040}, \href
  {http://adsabs.harvard.edu/abs/2006ApJ...642..752A} {642, 752}

\bibitem[\protect\citeauthoryear{{Battaglia}, {Helmi}, {Tolstoy}, {Irwin},
  {Hill}  \& {Jablonka}}{{Battaglia} et~al.}{2008}]{2008ApJ...681L..13B}
{Battaglia} G.,  {Helmi} A.,  {Tolstoy} E.,  {Irwin} M.,  {Hill} V.,
  {Jablonka} P.,  2008, \mn@doi [\apjl] {10.1086/590179}, \href
  {http://adsabs.harvard.edu/abs/2008ApJ...681L..13B} {681, L13}

\bibitem[\protect\citeauthoryear{{Baumgardt}, {Makino}, {Hut}, {McMillan}  \&
  {Portegies Zwart}}{{Baumgardt} et~al.}{2003}]{2003ApJ...589L..25B}
{Baumgardt} H.,  {Makino} J.,  {Hut} P.,  {McMillan} S.,   {Portegies Zwart}
  S.,  2003, \mn@doi [\apjl] {10.1086/375802}, \href
  {http://esoads.eso.org/abs/2003ApJ...589L..25B} {589, L25}

\bibitem[\protect\citeauthoryear{{Binney}}{{Binney}}{1980}]{1980MNRAS.190..873B}
{Binney} J.,  1980, \mn@doi [\mnras] {10.1093/mnras/190.4.873}, \href
  {http://esoads.eso.org/abs/1980MNRAS.190..873B} {190, 873}

\bibitem[\protect\citeauthoryear{{Binney} \& {Mamon}}{{Binney} \&
  {Mamon}}{1982}]{1982MNRAS.200..361B}
{Binney} J.,  {Mamon} G.~A.,  1982, \mn@doi [\mnras] {10.1093/mnras/200.2.361},
  \href {http://esoads.eso.org/abs/1982MNRAS.200..361B} {200, 361}

\bibitem[\protect\citeauthoryear{{Binney} \& {Tremaine}}{{Binney} \&
  {Tremaine}}{2008}]{1987gady.book.....B}
{Binney} J.,  {Tremaine} S.,  2008, {Galactic dynamics}.
Princeton, NJ, Princeton University Press, 2008, 747 p., \url
  {http://adsabs.harvard.edu/cgi-bin/nph-bib_query?bibcode=1987gady.book.....B&db_key=AST}

\bibitem[\protect\citeauthoryear{{Breddels} \& {Helmi}}{{Breddels} \&
  {Helmi}}{2013}]{2013A&A...558A..35B}
{Breddels} M.~A.,  {Helmi} A.,  2013, \mn@doi [\aap]
  {10.1051/0004-6361/201321606}, \href
  {http://adsabs.harvard.edu/abs/2013A%26A...558A..35B} {558, A35}

\bibitem[\protect\citeauthoryear{{Breddels} \& {Helmi}}{{Breddels} \&
  {Helmi}}{2014}]{2014ApJ...791L...3B}
{Breddels} M.~A.,  {Helmi} A.,  2014, \mn@doi [\apjl]
  {10.1088/2041-8205/791/1/L3}, \href
  {http://adsabs.harvard.edu/abs/2014ApJ...791L...3B} {791, L3}

\bibitem[\protect\citeauthoryear{{Breddels}, {Helmi}, {van den Bosch}, {van de
  Ven}  \& {Battaglia}}{{Breddels} et~al.}{2013}]{2013MNRAS.433.3173B}
{Breddels} M.~A.,  {Helmi} A.,  {van den Bosch} R.~C.~E.,  {van de Ven} G.,
  {Battaglia} G.,  2013, \mn@doi [\mnras] {10.1093/mnras/stt956}, \href
  {http://adsabs.harvard.edu/abs/2013MNRAS.433.3173B} {433, 3173}

\bibitem[\protect\citeauthoryear{{Cappellari}}{{Cappellari}}{2016}]{2016ARA&A..54..597C}
{Cappellari} M.,  2016, \mn@doi [\araa] {10.1146/annurev-astro-082214-122432},
  \href {http://esoads.eso.org/abs/2016ARA%26A..54..597C} {54, 597}

\bibitem[\protect\citeauthoryear{{Cappellari} et~al.,}{{Cappellari}
  et~al.}{2012}]{2012Natur.484..485C}
{Cappellari} M.,  et~al., 2012, \mn@doi [\nat] {10.1038/nature10972}, \href
  {http://esoads.eso.org/abs/2012Natur.484..485C} {484, 485}

\bibitem[\protect\citeauthoryear{{Ciotti} \& {Morganti}}{{Ciotti} \&
  {Morganti}}{2010}]{2010MNRAS.408.1070C}
{Ciotti} L.,  {Morganti} L.,  2010, \mn@doi [\mnras]
  {10.1111/j.1365-2966.2010.17184.x}, \href
  {http://esoads.eso.org/abs/2010MNRAS.408.1070C} {408, 1070}

\bibitem[\protect\citeauthoryear{{Coles}, {Read}  \& {Saha}}{{Coles}
  et~al.}{2014}]{2014MNRAS.445.2181C}
{Coles} J.~P.,  {Read} J.~I.,   {Saha} P.,  2014, \mn@doi [\mnras]
  {10.1093/mnras/stu1781}, \href
  {http://adsabs.harvard.edu/abs/2014MNRAS.445.2181C} {445, 2181}

\bibitem[\protect\citeauthoryear{{Cuddeford}}{{Cuddeford}}{1991}]{1991MNRAS.253..414C}
{Cuddeford} P.,  1991, \mnras, \href
  {http://adsabs.harvard.edu/cgi-bin/nph-bib_query?bibcode=1991MNRAS.253..414C&db_key=AST}
  {253, 414}

\bibitem[\protect\citeauthoryear{{Dehnen}}{{Dehnen}}{2009}]{2009MNRAS.395.1079D}
{Dehnen} W.,  2009, \mn@doi [\mnras] {10.1111/j.1365-2966.2009.14603.x}, \href
  {http://adsabs.harvard.edu/abs/2009MNRAS.395.1079D} {395, 1079}

\bibitem[\protect\citeauthoryear{{Eddington}}{{Eddington}}{1915}]{1915MNRAS..75..366E}
{Eddington} A.~S.,  1915, \mn@doi [\mnras] {10.1093/mnras/75.5.366}, \href
  {http://esoads.eso.org/abs/1915MNRAS..75..366E} {75, 366}

\bibitem[\protect\citeauthoryear{{Emsellem}, {Monnet}  \& {Bacon}}{{Emsellem}
  et~al.}{1994}]{1994A&A...285..723E}
{Emsellem} E.,  {Monnet} G.,   {Bacon} R.,  1994, \aap, \href
  {http://esoads.eso.org/abs/1994A%26A...285..723E} {285, 723}

\bibitem[\protect\citeauthoryear{{Famaey} \& {McGaugh}}{{Famaey} \&
  {McGaugh}}{2012}]{2012LRR....15...10F}
{Famaey} B.,  {McGaugh} S.~S.,  2012, \mn@doi [Living Reviews in Relativity]
  {10.12942/lrr-2012-10}, \href {http://esoads.eso.org/abs/2012LRR....15...10F}
  {15, 10}

\bibitem[\protect\citeauthoryear{{Foreman-Mackey}, {Hogg}, {Lang}  \&
  {Goodman}}{{Foreman-Mackey} et~al.}{2013}]{2013PASP..125..306F}
{Foreman-Mackey} D.,  {Hogg} D.~W.,  {Lang} D.,   {Goodman} J.,  2013, \mn@doi
  [\pasp] {10.1086/670067}, \href
  {http://adsabs.harvard.edu/abs/2013PASP..125..306F} {125, 306}

\bibitem[\protect\citeauthoryear{{Gebhardt}, {Rich}  \& {Ho}}{{Gebhardt}
  et~al.}{2005}]{2005ApJ...634.1093G}
{Gebhardt} K.,  {Rich} R.~M.,   {Ho} L.~C.,  2005, \mn@doi [\apj]
  {10.1086/497023}, \href {http://esoads.eso.org/abs/2005ApJ...634.1093G} {634,
  1093}

\bibitem[\protect\citeauthoryear{{Gieles} \& {Zocchi}}{{Gieles} \&
  {Zocchi}}{2015}]{2015MNRAS.454..576G}
{Gieles} M.,  {Zocchi} A.,  2015, \mn@doi [\mnras] {10.1093/mnras/stv1848},
  \href {http://esoads.eso.org/abs/2015MNRAS.454..576G} {454, 576}

\bibitem[\protect\citeauthoryear{{Gonz{\'a}les-Morales}, {Marsh},
  {Pe{\~n}arrubia}  \& {Ure{\~n}a-L{\'o}pez}}{{Gonz{\'a}les-Morales}
  et~al.}{2016}]{2016arXiv160905856G}
{Gonz{\'a}les-Morales} A.~X.,  {Marsh} D.~J.~E.,  {Pe{\~n}arrubia} J.,
  {Ure{\~n}a-L{\'o}pez} L.,  2016, preprint, \href
  {http://esoads.eso.org/abs/2016arXiv160905856G} {} (\mn@eprint {arXiv}
  {1609.05856})

\bibitem[\protect\citeauthoryear{{Gualandris}, {Read}, {Dehnen}  \&
  {Bortolas}}{{Gualandris} et~al.}{2017}]{2017MNRAS.464.2301G}
{Gualandris} A.,  {Read} J.~I.,  {Dehnen} W.,   {Bortolas} E.,  2017, \mn@doi
  [\mnras] {10.1093/mnras/stw2528}, \href
  {http://esoads.eso.org/abs/2017MNRAS.464.2301G} {464, 2301}

\bibitem[\protect\citeauthoryear{{Heggie} \& {Ramamani}}{{Heggie} \&
  {Ramamani}}{1995}]{1995MNRAS.272..317H}
{Heggie} D.~C.,  {Ramamani} N.,  1995, \mn@doi [\mnras]
  {10.1093/mnras/272.2.317}, \href
  {http://esoads.eso.org/abs/1995MNRAS.272..317H} {272, 317}

\bibitem[\protect\citeauthoryear{{Hernquist}}{{Hernquist}}{1990}]{1990ApJ...356..359H}
{Hernquist} L.,  1990, \apj, \href
  {http://adsabs.harvard.edu/cgi-bin/nph-bib_query?bibcode=1990ApJ...356..359H&db_key=AST}
  {356, 359}

\bibitem[\protect\citeauthoryear{{Hunt} \& {Kawata}}{{Hunt} \&
  {Kawata}}{2013}]{2013MNRAS.430.1928H}
{Hunt} J.~A.~S.,  {Kawata} D.,  2013, \mn@doi [\mnras] {10.1093/mnras/stt021},
  \href {http://adsabs.harvard.edu/abs/2013MNRAS.430.1928H} {430, 1928}

\bibitem[\protect\citeauthoryear{{Irwin} \& {Hatzidimitriou}}{{Irwin} \&
  {Hatzidimitriou}}{1995}]{1995MNRAS.277.1354I}
{Irwin} M.,  {Hatzidimitriou} D.,  1995, \mnras, \href
  {http://adsabs.harvard.edu/cgi-bin/nph-bib_query?bibcode=1995MNRAS.277.1354I&db_key=AST}
  {277, 1354}

\bibitem[\protect\citeauthoryear{{Jeans}}{{Jeans}}{1922}]{1922MNRAS..82..122J}
{Jeans} J.~H.,  1922, \mnras, \href
  {http://adsabs.harvard.edu/abs/1922MNRAS..82..122J} {82, 122}

\bibitem[\protect\citeauthoryear{{Kapteyn}}{{Kapteyn}}{1922}]{1922ApJ....55..302K}
{Kapteyn} J.~C.,  1922, \mn@doi [\apj] {10.1086/142670}, \href
  {http://adsabs.harvard.edu/abs/1922ApJ....55..302K} {55, 302}

\bibitem[\protect\citeauthoryear{{King}}{{King}}{1966}]{1966AJ.....71...64K}
{King} I.~R.,  1966, \mn@doi [\aj] {10.1086/109857}, \href
  {http://esoads.eso.org/abs/1966AJ.....71...64K} {71, 64}

\bibitem[\protect\citeauthoryear{{Kleyna}, {Wilkinson}, {Evans}  \&
  {Gilmore}}{{Kleyna} et~al.}{2001}]{KleynaEtal2001}
{Kleyna} J.~T.,  {Wilkinson} M.~I.,  {Evans} N.~W.,   {Gilmore} G.,  2001,
  \mn@doi [\apjl] {10.1086/338603}, 563, L115

\bibitem[\protect\citeauthoryear{{{\L}okas} \& {Mamon}}{{{\L}okas} \&
  {Mamon}}{2003}]{2003MNRAS.343..401L}
{{\L}okas} E.~L.,  {Mamon} G.~A.,  2003, \mn@doi [\mnras]
  {10.1046/j.1365-8711.2003.06684.x}, \href
  {http://adsabs.harvard.edu/abs/2003MNRAS.343..401L} {343, 401}

\bibitem[\protect\citeauthoryear{{L{\"u}ghausen}, {Famaey}  \&
  {Kroupa}}{{L{\"u}ghausen} et~al.}{2014}]{2014MNRAS.441.2497L}
{L{\"u}ghausen} F.,  {Famaey} B.,   {Kroupa} P.,  2014, \mn@doi [\mnras]
  {10.1093/mnras/stu757}, \href {http://esoads.eso.org/abs/2014MNRAS.441.2497L}
  {441, 2497}

\bibitem[\protect\citeauthoryear{{Magorrian} et~al.,}{{Magorrian}
  et~al.}{1998}]{MagorrianEtal1998}
{Magorrian} J.,  et~al., 1998, \aj, 115, 2285

\bibitem[\protect\citeauthoryear{{Mamon} \& {{\L}okas}}{{Mamon} \&
  {{\L}okas}}{2005}]{2005MNRAS.363..705M}
{Mamon} G.~A.,  {{\L}okas} E.~L.,  2005, \mn@doi [\mnras]
  {10.1111/j.1365-2966.2005.09400.x}, \href
  {http://esoads.eso.org/abs/2005MNRAS.363..705M} {363, 705}

\bibitem[\protect\citeauthoryear{{Mamon}, {Biviano}  \& {Bou{\'e}}}{{Mamon}
  et~al.}{2013}]{2013MNRAS.429.3079M}
{Mamon} G.~A.,  {Biviano} A.,   {Bou{\'e}} G.,  2013, \mn@doi [\mnras]
  {10.1093/mnras/sts565}, \href {http://esoads.eso.org/abs/2013MNRAS.429.3079M}
  {429, 3079}

\bibitem[\protect\citeauthoryear{{Martin}, {de Jong}  \& {Rix}}{{Martin}
  et~al.}{2008}]{2008ApJ...684.1075M}
{Martin} N.~F.,  {de Jong} J.~T.~A.,   {Rix} H.-W.,  2008, \mn@doi [\apj]
  {10.1086/590336}, \href {http://adsabs.harvard.edu/abs/2008ApJ...684.1075M}
  {684, 1075}

\bibitem[\protect\citeauthoryear{{McGaugh}}{{McGaugh}}{2016}]{2016ApJ...832L...8M}
{McGaugh} S.~S.,  2016, \mn@doi [\apjl] {10.3847/2041-8205/832/1/L8}, \href
  {http://esoads.eso.org/abs/2016ApJ...832L...8M} {832, L8}

\bibitem[\protect\citeauthoryear{{Merrifield} \& {Kent}}{{Merrifield} \&
  {Kent}}{1990}]{1990AJ.....99.1548M}
{Merrifield} M.~R.,  {Kent} S.~M.,  1990, \mn@doi [\aj] {10.1086/115438}, \href
  {http://esoads.eso.org/abs/1990AJ.....99.1548M} {99, 1548}

\bibitem[\protect\citeauthoryear{{Merritt}}{{Merritt}}{1985}]{1985MNRAS.214P..25M}
{Merritt} D.,  1985, \mnras, \href
  {http://adsabs.harvard.edu/cgi-bin/nph-bib_query?bibcode=1985MNRAS.214P..25M&db_key=AST}
  {214, 25P}

\bibitem[\protect\citeauthoryear{{Miller-Jones} et~al.,}{{Miller-Jones}
  et~al.}{2012}]{2012ApJ...755L...1M}
{Miller-Jones} J.~C.~A.,  et~al., 2012, \mn@doi [\apjl]
  {10.1088/2041-8205/755/1/L1}, \href
  {http://esoads.eso.org/abs/2012ApJ...755L...1M} {755, L1}

\bibitem[\protect\citeauthoryear{{Napolitano}, {Pota}, {Romanowsky}, {Forbes},
  {Brodie}  \& {Foster}}{{Napolitano} et~al.}{2014}]{2014MNRAS.439..659N}
{Napolitano} N.~R.,  {Pota} V.,  {Romanowsky} A.~J.,  {Forbes} D.~A.,  {Brodie}
  J.~P.,   {Foster} C.,  2014, \mn@doi [\mnras] {10.1093/mnras/stt2484}, \href
  {http://esoads.eso.org/abs/2014MNRAS.439..659N} {439, 659}

\bibitem[\protect\citeauthoryear{{Navarro}, {Eke}  \& {Frenk}}{{Navarro}
  et~al.}{1996a}]{1996MNRAS.283L..72N}
{Navarro} J.~F.,  {Eke} V.~R.,   {Frenk} C.~S.,  1996a, \mnras, 283, L72

\bibitem[\protect\citeauthoryear{{Navarro}, {Frenk}  \& {White}}{{Navarro}
  et~al.}{1996b}]{1996ApJ...462..563N}
{Navarro} J.~F.,  {Frenk} C.~S.,   {White} S.~D.~M.,  1996b, \apj, 462, 563

\bibitem[\protect\citeauthoryear{Newville, Stensitzki, Allen  \&
  Ingargiola}{Newville et~al.}{2014}]{newville_2014_11813}
Newville M.,  Stensitzki T.,  Allen D.~B.,   Ingargiola A.,  2014, {LMFIT:
  Non-Linear Least-Square Minimization and Curve-Fitting for Python¶},
  \mn@doi{10.5281/zenodo.11813}, \url {https://doi.org/10.5281/zenodo.11813}

\bibitem[\protect\citeauthoryear{{Noyola}, {Gebhardt}  \& {Bergmann}}{{Noyola}
  et~al.}{2008}]{2008ApJ...676.1008N}
{Noyola} E.,  {Gebhardt} K.,   {Bergmann} M.,  2008, \mn@doi [\apj]
  {10.1086/529002}, \href {http://esoads.eso.org/abs/2008ApJ...676.1008N} {676,
  1008}

\bibitem[\protect\citeauthoryear{{Noyola}, {Gebhardt}, {Kissler-Patig},
  {L{\"u}tzgendorf}, {Jalali}, {de Zeeuw}  \& {Baumgardt}}{{Noyola}
  et~al.}{2010}]{2010ApJ...719L..60N}
{Noyola} E.,  {Gebhardt} K.,  {Kissler-Patig} M.,  {L{\"u}tzgendorf} N.,
  {Jalali} B.,  {de Zeeuw} P.~T.,   {Baumgardt} H.,  2010, \mn@doi [\apjl]
  {10.1088/2041-8205/719/1/L60}, \href
  {http://esoads.eso.org/abs/2010ApJ...719L..60N} {719, L60}

\bibitem[\protect\citeauthoryear{{Osipkov}}{{Osipkov}}{1979}]{1979PAZh....5...77O}
{Osipkov} L.~P.,  1979, Pis ma Astronomicheskii Zhurnal, \href
  {http://adsabs.harvard.edu/cgi-bin/nph-bib_query?bibcode=1979PAZh....5...77O&db_key=AST}
  {5, 77}

\bibitem[\protect\citeauthoryear{{Peuten}, {Zocchi}, {Gieles}, {Gualandris}  \&
  {H{\'e}nault-Brunet}}{{Peuten} et~al.}{2016}]{2016MNRAS.462.2333P}
{Peuten} M.,  {Zocchi} A.,  {Gieles} M.,  {Gualandris} A.,
  {H{\'e}nault-Brunet} V.,  2016, \mn@doi [\mnras] {10.1093/mnras/stw1726},
  \href {http://esoads.eso.org/abs/2016MNRAS.462.2333P} {462, 2333}

\bibitem[\protect\citeauthoryear{{Plummer}}{{Plummer}}{1911}]{1911MNRAS..71..460P}
{Plummer} H.~C.,  1911, \mnras, \href
  {http://adsabs.harvard.edu/cgi-bin/nph-bib_query?bibcode=1911MNRAS..71..460P&db_key=AST}
  {71, 460}

\bibitem[\protect\citeauthoryear{{Pontzen} \& {Governato}}{{Pontzen} \&
  {Governato}}{2012}]{2012MNRAS.421.3464P}
{Pontzen} A.,  {Governato} F.,  2012, \mn@doi [\mnras]
  {10.1111/j.1365-2966.2012.20571.x}, \href
  {http://adsabs.harvard.edu/abs/2012MNRAS.421.3464P} {421, 3464}

\bibitem[\protect\citeauthoryear{{Pryor} \& {Meylan}}{{Pryor} \&
  {Meylan}}{1993}]{1993ASPC...50..357P}
{Pryor} C.,  {Meylan} G.,  1993, in Astronomical Society of the Pacific
  Conference Series. pp 357--+

\bibitem[\protect\citeauthoryear{{Read} \& {Gilmore}}{{Read} \&
  {Gilmore}}{2005}]{2005MNRAS.356..107R}
{Read} J.~I.,  {Gilmore} G.,  2005, \mnras, \href
  {http://adsabs.harvard.edu/cgi-bin/nph-bib_query?bibcode=2005MNRAS.356..107R&db_key=AST}
  {356, 107}

\bibitem[\protect\citeauthoryear{{Read}, {Wilkinson}, {Evans}, {Gilmore}  \&
  {Kleyna}}{{Read} et~al.}{2006}]{2006MNRAS.tmp..153R}
{Read} J.~I.,  {Wilkinson} M.~I.,  {Evans} N.~W.,  {Gilmore} G.,   {Kleyna}
  J.~T.,  2006, \mn@doi [\mnras] {10.1111/j.1365-2966.2005.09959.x}, \href
  {http://adsabs.harvard.edu/cgi-bin/nph-bib_query?bibcode=2006MNRAS.367..387R&db_key=AST}
  {367, 387}

\bibitem[\protect\citeauthoryear{{Richardson} \& {Fairbairn}}{{Richardson} \&
  {Fairbairn}}{2013}]{2013MNRAS.432.3361R}
{Richardson} T.,  {Fairbairn} M.,  2013, \mn@doi [\mnras]
  {10.1093/mnras/stt686}, \href {http://esoads.eso.org/abs/2013MNRAS.432.3361R}
  {432, 3361}

\bibitem[\protect\citeauthoryear{{Richardson} \& {Fairbairn}}{{Richardson} \&
  {Fairbairn}}{2014}]{2014MNRAS.441.1584R}
{Richardson} T.,  {Fairbairn} M.,  2014, \mn@doi [\mnras]
  {10.1093/mnras/stu691}, \href
  {http://adsabs.harvard.edu/abs/2014MNRAS.441.1584R} {441, 1584}

\bibitem[\protect\citeauthoryear{{Rojas-Ni{\~n}o}, {Read}, {Aguilar}  \&
  {Delorme}}{{Rojas-Ni{\~n}o} et~al.}{2016}]{2016MNRAS.459.3349R}
{Rojas-Ni{\~n}o} A.,  {Read} J.~I.,  {Aguilar} L.,   {Delorme} M.,  2016,
  \mn@doi [\mnras] {10.1093/mnras/stw846}, \href
  {http://esoads.eso.org/abs/2016MNRAS.459.3349R} {459, 3349}

\bibitem[\protect\citeauthoryear{{Romanowsky}, {Douglas}, {Arnaboldi},
  {Kuijken}, {Merrifield}, {Napolitano}, {Capaccioli}  \&
  {Freeman}}{{Romanowsky} et~al.}{2003}]{2003Sci...301.1696R}
{Romanowsky} A.~J.,  {Douglas} N.~G.,  {Arnaboldi} M.,  {Kuijken} K.,
  {Merrifield} M.~R.,  {Napolitano} N.~R.,  {Capaccioli} M.,   {Freeman} K.~C.,
   2003, Science, \href
  {http://adsabs.harvard.edu/cgi-bin/nph-bib_query?bibcode=2003Sci...301.1696R&db_key=AST}
  {301, 1696}

\bibitem[\protect\citeauthoryear{{Saglia}, {Kronawitter}, {Gerhard}  \&
  {Bender}}{{Saglia} et~al.}{2000}]{2000AJ....119..153S}
{Saglia} R.~P.,  {Kronawitter} A.,  {Gerhard} O.,   {Bender} R.,  2000, \mn@doi
  [\aj] {10.1086/301153}, \href {http://esoads.eso.org/abs/2000AJ....119..153S}
  {119, 153}

\bibitem[\protect\citeauthoryear{{Sanders} \& {Evans}}{{Sanders} \&
  {Evans}}{2016}]{2016ApJ...830L..26S}
{Sanders} J.~L.,  {Evans} N.~W.,  2016, \mn@doi [\apjl]
  {10.3847/2041-8205/830/2/L26}, \href
  {http://esoads.eso.org/abs/2016ApJ...830L..26S} {830, L26}

\bibitem[\protect\citeauthoryear{{Schwarzschild}}{{Schwarzschild}}{1979}]{1979ApJ...232..236S}
{Schwarzschild} M.,  1979, \mn@doi [\apj] {10.1086/157282}, \href
  {http://adsabs.harvard.edu/cgi-bin/nph-bib_query?bibcode=1979ApJ...232..236S&db_key=AST}
  {232, 236}

\bibitem[\protect\citeauthoryear{{Shanahan} \& {Gieles}}{{Shanahan} \&
  {Gieles}}{2015}]{2015MNRAS.448L..94S}
{Shanahan} R.~L.,  {Gieles} M.,  2015, \mn@doi [\mnras]
  {10.1093/mnrasl/slu205}, \href
  {http://esoads.eso.org/abs/2015MNRAS.448L..94S} {448, L94}

\bibitem[\protect\citeauthoryear{{Sollima}, {Bellazzini}  \& {Lee}}{{Sollima}
  et~al.}{2012}]{2012ApJ...755..156S}
{Sollima} A.,  {Bellazzini} M.,   {Lee} J.-W.,  2012, \mn@doi [\apj]
  {10.1088/0004-637X/755/2/156}, \href
  {http://esoads.eso.org/abs/2012ApJ...755..156S} {755, 156}

\bibitem[\protect\citeauthoryear{{Strigari}, {Bullock}  \&
  {Kaplinghat}}{{Strigari} et~al.}{2007}]{2007ApJ...657L...1S}
{Strigari} L.~E.,  {Bullock} J.~S.,   {Kaplinghat} M.,  2007, \mn@doi [\apjl]
  {10.1086/512976}, \href {http://esoads.eso.org/abs/2007ApJ...657L...1S} {657,
  L1}

\bibitem[\protect\citeauthoryear{{Syer} \& {Tremaine}}{{Syer} \&
  {Tremaine}}{1996}]{1996MNRAS.282..223S}
{Syer} D.,  {Tremaine} S.,  1996, \mnras, \href
  {http://adsabs.harvard.edu/cgi-bin/nph-bib_query?bibcode=1996MNRAS.282..223S&db_key=AST}
  {282, 223}

\bibitem[\protect\citeauthoryear{{Walker} \& {Pe{\~n}arrubia}}{{Walker} \&
  {Pe{\~n}arrubia}}{2011}]{2011ApJ...742...20W}
{Walker} M.~G.,  {Pe{\~n}arrubia} J.,  2011, \mn@doi [\apj]
  {10.1088/0004-637X/742/1/20}, \href
  {http://adsabs.harvard.edu/abs/2011ApJ...742...20W} {742, 20}

\bibitem[\protect\citeauthoryear{{Walker}, {Mateo}  \& {Olszewski}}{{Walker}
  et~al.}{2009a}]{2009AJ....137.3100W}
{Walker} M.~G.,  {Mateo} M.,   {Olszewski} E.~W.,  2009a, \mn@doi [\aj]
  {10.1088/0004-6256/137/2/3100}, \href
  {http://adsabs.harvard.edu/abs/2009AJ....137.3100W} {137, 3100}

\bibitem[\protect\citeauthoryear{{Walker}, {Mateo}, {Olszewski},
  {Pe{\~n}arrubia}, {Wyn Evans}  \& {Gilmore}}{{Walker}
  et~al.}{2009b}]{2009ApJ...704.1274W}
{Walker} M.~G.,  {Mateo} M.,  {Olszewski} E.~W.,  {Pe{\~n}arrubia} J.,  {Wyn
  Evans} N.,   {Gilmore} G.,  2009b, \mn@doi [\apj]
  {10.1088/0004-637X/704/2/1274}, \href
  {http://adsabs.harvard.edu/abs/2009ApJ...704.1274W} {704, 1274}

\bibitem[\protect\citeauthoryear{{Watkins}, {van de Ven}, {den Brok}  \& {van
  den Bosch}}{{Watkins} et~al.}{2013}]{2013MNRAS.436.2598W}
{Watkins} L.~L.,  {van de Ven} G.,  {den Brok} M.,   {van den Bosch} R.~C.~E.,
  2013, \mn@doi [\mnras] {10.1093/mnras/stt1756}, \href
  {http://esoads.eso.org/abs/2013MNRAS.436.2598W} {436, 2598}

\bibitem[\protect\citeauthoryear{{Wilkinson}, {Kleyna}, {Evans}  \&
  {Gilmore}}{{Wilkinson} et~al.}{2002}]{2002MNRAS.330..778W}
{Wilkinson} M.~I.,  {Kleyna} J.,  {Evans} N.~W.,   {Gilmore} G.,  2002, \mn@doi
  [\mnras] {10.1046/j.1365-8711.2002.05154.x}, \href
  {http://adsabs.harvard.edu/abs/2002MNRAS.330..778W} {330, 778}

\bibitem[\protect\citeauthoryear{{Wojtak}, {{\L}okas}, {Mamon}  \&
  {Gottl{\"o}ber}}{{Wojtak} et~al.}{2009}]{2009MNRAS.399..812W}
{Wojtak} R.,  {{\L}okas} E.~L.,  {Mamon} G.~A.,   {Gottl{\"o}ber} S.,  2009,
  \mn@doi [\mnras] {10.1111/j.1365-2966.2009.15312.x}, \href
  {http://esoads.eso.org/abs/2009MNRAS.399..812W} {399, 812}

\bibitem[\protect\citeauthoryear{{Wolf}, {Martinez}, {Bullock}, {Kaplinghat},
  {Geha}, {Mu{\~n}oz}, {Simon}  \& {Avedo}}{{Wolf}
  et~al.}{2010}]{2010MNRAS.406.1220W}
{Wolf} J.,  {Martinez} G.~D.,  {Bullock} J.~S.,  {Kaplinghat} M.,  {Geha} M.,
  {Mu{\~n}oz} R.~R.,  {Simon} J.~D.,   {Avedo} F.~F.,  2010, \mn@doi [\mnras]
  {10.1111/j.1365-2966.2010.16753.x}, \href
  {http://adsabs.harvard.edu/abs/2010MNRAS.406.1220W} {406, 1220}

\bibitem[\protect\citeauthoryear{{Young}, {Williams}  \& {Hjorth}}{{Young}
  et~al.}{2016}]{2016JCAP...05..010Y}
{Young} A.~M.,  {Williams} L.~L.~R.,   {Hjorth} J.,  2016, \mn@doi [\jcap]
  {10.1088/1475-7516/2016/05/010}, \href
  {http://adsabs.harvard.edu/abs/2016JCAP...05..010Y} {5, 010}

\bibitem[\protect\citeauthoryear{{Zhao}}{{Zhao}}{1996}]{1996MNRAS.278..488Z}
{Zhao} H.,  1996, \mnras, \href
  {http://adsabs.harvard.edu/cgi-bin/nph-bib_query?bibcode=1996MNRAS.278..488Z&db_key=AST}
  {278, 488}

\bibitem[\protect\citeauthoryear{{Zocchi}, {Gieles}  \&
  {H{\'e}nault-Brunet}}{{Zocchi} et~al.}{2016}]{2016IAUS..312..197Z}
{Zocchi} A.,  {Gieles} M.,   {H{\'e}nault-Brunet} V.,  2016, in {Meiron} Y.,
  {Li} S.,  {Liu} F.-K.,   {Spurzem} R.,  eds,  IAU Symposium Vol. 312, Star
  Clusters and Black Holes in Galaxies across Cosmic Time. pp 197--200
  (\mn@eprint {arXiv} {1501.05262}), \mn@doi{10.1017/S1743921315007802}

\bibitem[\protect\citeauthoryear{{Zwicky}}{{Zwicky}}{1937}]{1937ApJ....86..217Z}
{Zwicky} F.,  1937, \apj, 86, 217

\bibitem[\protect\citeauthoryear{{de Lorenzi}, {Debattista}, {Gerhard}  \&
  {Sambhus}}{{de Lorenzi} et~al.}{2007}]{2007MNRAS.376...71D}
{de Lorenzi} F.,  {Debattista} V.~P.,  {Gerhard} O.,   {Sambhus} N.,  2007,
  \mn@doi [\mnras] {10.1111/j.1365-2966.2007.11434.x}, \href
  {http://adsabs.harvard.edu/abs/2007MNRAS.376...71D} {376, 71}

\bibitem[\protect\citeauthoryear{{de Lorenzi} et~al.,}{{de Lorenzi}
  et~al.}{2009}]{2009MNRAS.395...76D}
{de Lorenzi} F.,  et~al., 2009, \mn@doi [\mnras]
  {10.1111/j.1365-2966.2009.14553.x}, \href
  {http://esoads.eso.org/abs/2009MNRAS.395...76D} {395, 76}

\bibitem[\protect\citeauthoryear{{van den Bosch} \& {de Zeeuw}}{{van den Bosch}
  \& {de Zeeuw}}{2010}]{2010MNRAS.401.1770V}
{van den Bosch} R.~C.~E.,  {de Zeeuw} P.~T.,  2010, \mn@doi [\mnras]
  {10.1111/j.1365-2966.2009.15832.x}, \href
  {http://esoads.eso.org/abs/2010MNRAS.401.1770V} {401, 1770}

\bibitem[\protect\citeauthoryear{{van den Bosch}, {van de Ven}, {Verolme},
  {Cappellari}  \& {de Zeeuw}}{{van den Bosch}
  et~al.}{2008}]{2008MNRAS.385..647V}
{van den Bosch} R.~C.~E.,  {van de Ven} G.,  {Verolme} E.~K.,  {Cappellari} M.,
    {de Zeeuw} P.~T.,  2008, \mn@doi [\mnras]
  {10.1111/j.1365-2966.2008.12874.x}, \href
  {http://adsabs.harvard.edu/abs/2008MNRAS.385..647V} {385, 647}

\bibitem[\protect\citeauthoryear{{van der Marel}}{{van der
  Marel}}{1994}]{1994MNRAS.270..271V}
{van der Marel} R.~P.,  1994, \mn@doi [\mnras] {10.1093/mnras/270.2.271}, \href
  {http://esoads.eso.org/abs/1994MNRAS.270..271V} {270, 271}

\bibitem[\protect\citeauthoryear{{van der Marel} \& {Anderson}}{{van der Marel}
  \& {Anderson}}{2010}]{2010ApJ...710.1063V}
{van der Marel} R.~P.,  {Anderson} J.,  2010, \mn@doi [\apj]
  {10.1088/0004-637X/710/2/1063}, \href
  {http://esoads.eso.org/abs/2010ApJ...710.1063V} {710, 1063}

\bibitem[\protect\citeauthoryear{{van der Marel} \& {Franx}}{{van der Marel} \&
  {Franx}}{1993}]{1993ApJ...407..525V}
{van der Marel} R.~P.,  {Franx} M.,  1993, \mn@doi [\apj] {10.1086/172534},
  \href {http://esoads.eso.org/abs/1993ApJ...407..525V} {407, 525}

\makeatother
\end{thebibliography}

\end{document}